\documentclass[10pt,a4paper]{article}
\usepackage{geometry}
\usepackage[utf8]{inputenc}

\usepackage{multicol}
\usepackage{diagbox}
\usepackage{caption}
\usepackage{subcaption}
\usepackage{jcappub}
\usepackage{bm}
\usepackage{epsfig}
\usepackage{bbold}
\usepackage{graphicx}
\usepackage{amsmath}
\usepackage{amssymb}
\usepackage{amsbsy}
\usepackage{color}
\usepackage{slashed}
\usepackage{afterpage}
\usepackage{psfrag}
\usepackage[noabbrev]{cleveref}
\usepackage{dsfont}
\usepackage{enumerate}
\usepackage[shortlabels]{enumitem}
\usepackage[utf8]{inputenc}
\usepackage{amsmath}
\usepackage[dvipsnames]{xcolor}
\usepackage{graphicx}
\usepackage{braket}
\usepackage[most]{tcolorbox}
\usepackage{empheq}

\def\parfrac#1#2{{\left(\frac{#1}{#2}\right) }}

\title{\begin{center}
		Axionic defects in the CMB: \\ birefringence and gravitational waves
\end{center}}

\vfill
\author[a]{Ricardo Z. Ferreira,}
\author[b,c]{Silvia Gasparotto,}
\author[d]{Takashi Hiramatsu,} 
\author[e]{Ippei Obata}
\author{and}
\author[c]{Oriol Pujolàs}

\affiliation[a]{CFisUC, Department of Physics, University of Coimbra, P-3004 - 516 Coimbra, Portugal}
\affiliation[b]{Grup de F\'{i}sica Te\`{o}rica, Departament de F\'{i}sica, Universitat Aut\`{o}noma de Barcelona, 08193 Bellaterra (Barcelona), Spain}{}
\affiliation[c]{Institut de Física d’Altes Energies (IFAE) and Barcelona Institute of Science and Technology (BIST), Campus UAB, 08193 Bellaterra, Barcelona, Spain}
\affiliation[d]{Department of Physics, Rikkyo University, Toshima, Tokyo 171-8501, Japan}
\affiliation[e]{Kavli Institute for the Physics and Mathematics of the Universe (WPI),
	University of Tokyo, Kashiwa, Chiba 277-8583, Japan}

\emailAdd{rzferreira@uc.pt}
\emailAdd{sgasparotto@ifae.es}
\emailAdd{hiramatz@rikkyo.ac.jp}
\emailAdd{ippei.obata@ipmu.jp}
\emailAdd{pujolas@ifae.es}

\makeatletter
\gdef\@fpheader{}
\makeatother

\abstract{
	The evidence for a non-vanishing isotropic cosmic birefringence in recent analyses of the CMB data provides a tantalizing hint for new physics.
	Domain wall (DW) networks have recently been shown to generate an isotropic birefringence signal in the ballpark of the measured value when coupled to photons.
 In this work, we explore the axionic defects hypothesis in more detail and extending previous results to annihilating and late-forming networks, and by pointing out other smoking-gun signatures of the network in the CMB spectrum such as the anisotropic birefringent spectrum and B-modes.
 We also argue that the presence of cosmic strings in the network does not hinder a large isotropic birefringence signal because of an intrinsic environmental contribution coming from low redshifts thus leaving open the possibility that axionic defects can explain the signal. 
 Regarding the remaining CMB signatures,  with the help of dedicated 3D numerical simulations of DW networks, that we took as a proxy for the axionic defects, we show how the anisotropic birefringence spectrum combined with a tomographic approach can be used to infer the formation and annihilation time of the network.   
  From the numerical simulations, we also computed the spectrum of gravitational waves (GWs) generated by the network in the post-recombination epoch 
and use previous searches for stochastic GW backgrounds in the CMB to derive for the first time a bound on the tension and abundance of networks with DWs that annihilate after recombination. Our bounds extend to the case where the network survives until the present time and improve over previous bounds by roughly one order of magnitude. Finally, we show the interesting prospects for detecting B-modes of DW origin with future CMB experiments.
}

\begin{document}
\hfill{\small RUP-23-27}
    	\maketitle

\section{Introduction}

Recent joint analyses of  the {\it Planck} and WMAP {\it EB} power spectra have measured an isotropic rotation of the Cosmic Microwave Background (CMB) polarization by an angle $\beta=0.342 \pm^{0.094}_{0.091} \deg$ at 68\% CL thus excluding the null hyphotesis  at $3.6 \sigma$ significance \cite{Eskilt:2022cff}. 
This effect, also known as a cosmic birefringence \cite{Carroll:1989vb, PhysRevD.43.3789, Harari:1992ea,Lue:1998mq}, has been probed in the past \cite{Feng:2006dp,QUaD:2008ado,WMAP:2010qai,Planck:2016soo}, but its detection has been strongly limited by the degeneracy with systematic errors in the instrumental calibration of the polarization angle. Recently, this degeneracy was broken by the following methodology:
miscalibration affects all electromagnetic sources, including galactic foregrounds, while a cosmological birefringence is not expected to rotate the galactic contribution significantly, given its relative proximity to the detector, thus allowing for a separation between the cosmological and the galactic foregound miscalibration components \cite{Minami:2019ruj,Minami:2020xfg,Minami:2020fin,Minami:2020odp,Diego-Palazuelos:2022dsq,Eskilt:2022wav,Eskilt:2023ndm}. 
The lack of a full understanding of the foreground contributions to the \textit{EB} cross-correlation still leaves some uncertainty on the cosmological origin of the effect \cite{Clark:2021kze,Diego-Palazuelos:2022cnh}, but the tantalizing measurement and the possibility that future CMB experiments will greatly improve the calibration accuracy \cite{Monelli:2022pru,Jost:2022oab} strongly motivates a search for the cosmological mechanisms that could be responsible for the cosmic birefringence. 

On the theory side, the situation is quite interesting and rather simple; to rotate the polarization axis of the CMB photons it is sufficient to have a time-dependent (pseudo-) scalar field $\phi$ coupled to photons via the electromagnetic anomaly term  \cite{Carroll:1998zi,Lue:1998mq} (see \cite{Komatsu:2022nvu} for a review), of the form
$$
c_\gamma \frac{\alpha_\text{em}}{8\pi}\,  \frac{\phi}{f} \, F_{\mu \nu}\tilde{F}^{\mu \nu}
$$ 
that is typically associated with axion-like particles \cite{Arvanitaki:2009fg}.
Any space/time dependent background in $\phi$ generates birefringence, with an amount determined by the variation of $\phi$ from the last scattering surface until today. 

Among the different ideas of explaining the birefringence measurement, two implementations have been singled out. One includes an ultra-light axion, that can be (early) dark energy or a subdominant component of dark matter, and that couples to photons as above \cite{Pospelov:2008gg,Finelli:2008jv,Panda:2010uq,Lee:2013mqa,Zhao:2014yna,Liu:2016dcg,Sigl:2018fba,Fedderke:2019ajk,Fujita:2020aqt,Fujita:2020ecn,Choi:2021aze,Obata:2021nql,Alvey:2021hjp,Nakatsuka:2022epj,Gasparotto:2022uqo,Murai:2022zur,Eskilt:2023nxm,Yin:2023srb,Gasparotto:2023psh}.
The second possibility is that the isotropic birefringence is caused by a network of cosmic domain walls (DWs) \cite{Takahashi:2020tqv,Kitajima:2022jzz,Gonzalez:2022mcx,Kitajima:2023kzu}  whose microphysical constituent couples to photons.  DWs are field configurations that result from the spontaneous breaking of a discrete symmetry and that connect between different (quasi-) degenerate minima. The polarization of a photon that couples to the wall will then be rotated by a sizeable amount when crossing it.  
In this case, the scalar field can have a larger mass than in the first possibility thus greatly enlarging the range of parameters that can explain the signal.
A potential explanation within the framework of low-energy effective field theory of Standard Model (SMEFT) is discussed in Ref. \cite{Nakai:2023zdr} concluding that no operator SMEFT could realize the measured isotropic signal, which further hints towards the need of light (axion-like) degrees of freedom.

The DW network interpretation of the isotropic birefringence hinges on two crucial ingredients: i) the coupling to the anomaly (with strength controlled by $1/f$) and ii) the DWs, which provide the space/time dependent background for $\phi$ with large `excursions', $\Delta \phi$, the difference in field space between the different vacua. For axion-like particles (axions, for short) $\Delta \phi$ is typically set by $f$, and this turns out to give the right ballpark to explain the isotropic birefringence \cite{Takahashi:2020tqv}, independently of the value of $f$, since this factors out after all. 
Overall, this raises an added interest in the axionic case as they are singled out as a simple explanation for isotropic birefringence.

Motivated by this, this work aims to explore in more detail the connection between the isotropic birefringence and the networks of axionic defects, and to identify additional CMB signatures of such connection. 
In particular, we shall develop three different aspects. First, the genericity of the isotropic birefringence signal in axionic models with defects. 
Second, explore in more detail the connection with anisotropic birefringence observables, aiming at the derivation of additional CMB signatures of the network that can distinguish this model from other interpretations of the signal. 
Last, we will focus on networks that include DWs and discuss one probe that is independent of the tension of the DW, the spectrum of gravitational waves generated by the network. 

\begin{figure}[t]
    \centering
\includegraphics[width=0.45\linewidth]{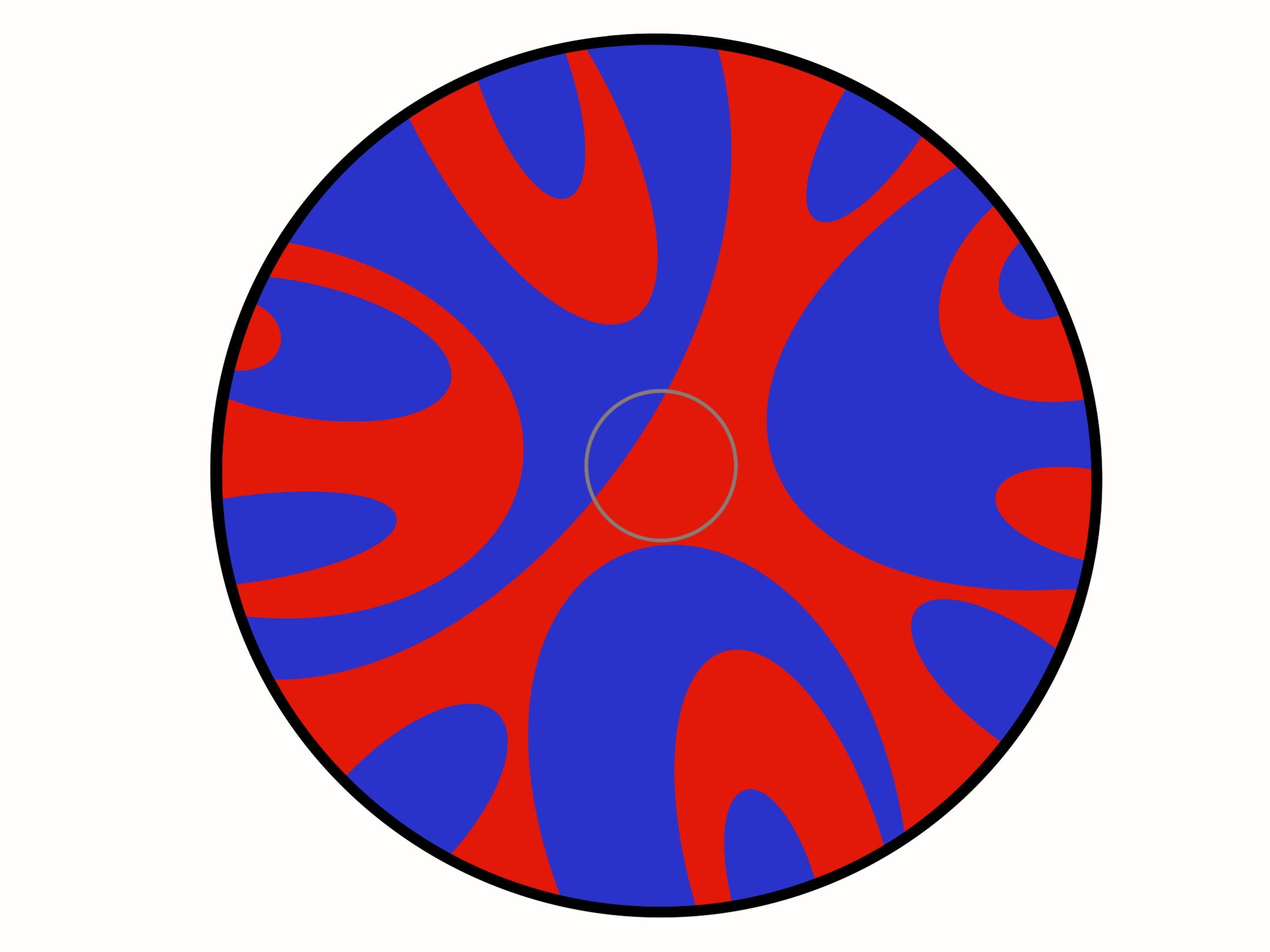}
    \caption{ 
    The projection in the sky of the DW network evolution can be represented in a ``celestial plot'', showing the configuration of the network (of the scalar field) as a function of time and the angle in the sky. Different colours correspond to different vacua.
    One can normalize the radial distance to be the redshift, $z$, so that the center of the disk is $z=0$ (here, now) and the boundary to, say, the CMB last scattering surface.
    The plot sketches how the evolution looks like in a $Z_2$ model with (2 degenerate vacua) during a scaling regime. At all times there must be roughly as many color changes (DWs) as the number of Hubble patches. The gray circle represents our local universe which can be taken as $z\simeq 1$.  
    }
    \label{fig:sketchZ2}
\end{figure}

\begin{figure}[t]
    \centering
\includegraphics[width=0.35\linewidth]{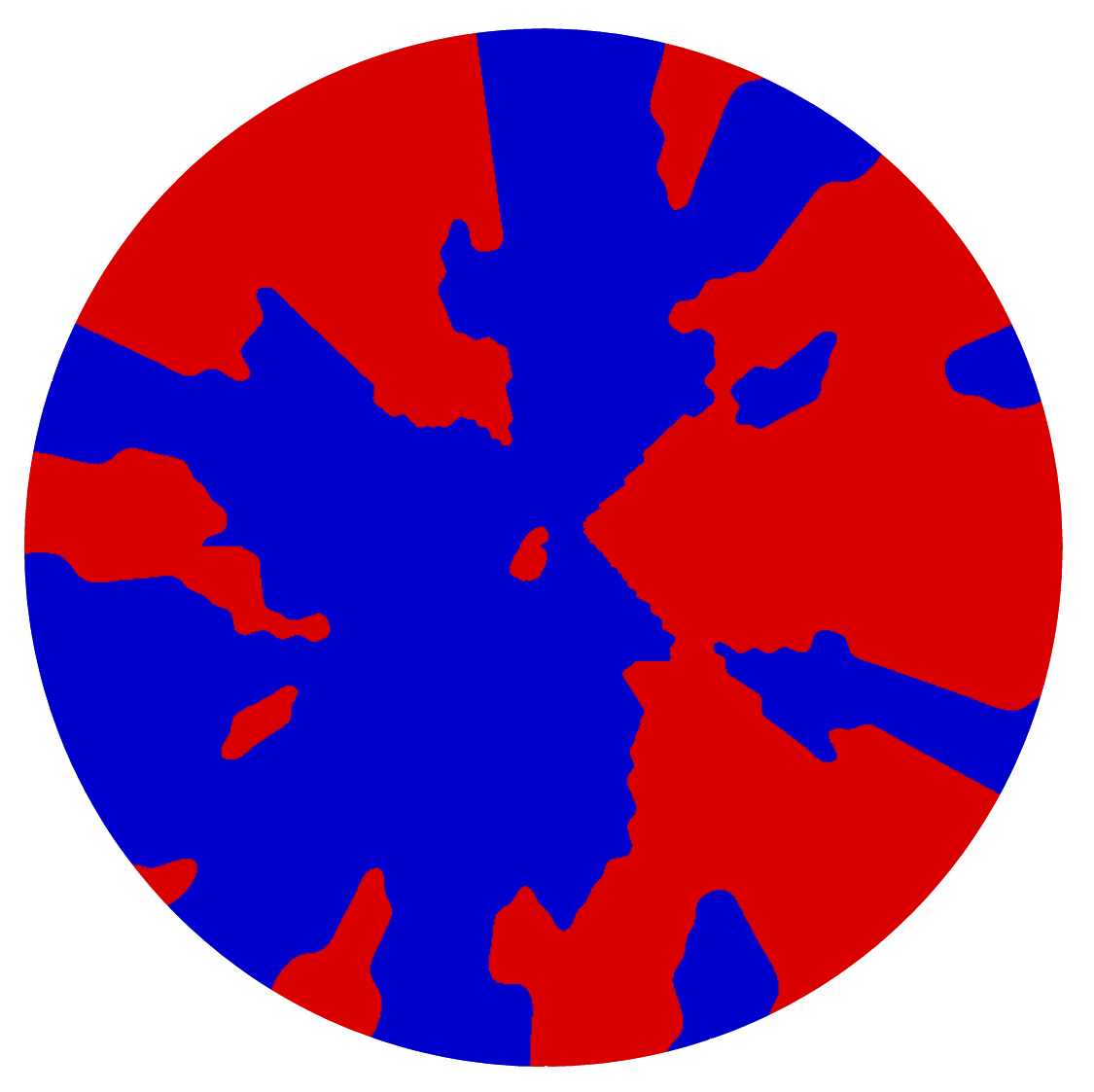}
    \caption{
 Same as Fig. 1 but with data obtained from one of our numerical simulations in a $Z_2$ model in $3+1$ dimensions with 2 degenerate vacua, that is without bias. The plot is obtained by evaluating (a spatial slice of) the scalar field in the past light cone and coloring the different vacua. 
    The appearance of 'islands' is due to the $3+1$ nature of the simulation: DWs can cross and then exit the chosen slice. Scaling is reached an some intermediate time.
    At all redshifts $z$, there are statistically equal numbers of Hubble patches of either color.  
   }
    \label{fig: past light cone}
\end{figure}

To see a first glimpse that topological defects may lead to isotropic birefringence, start with the DW case and assume that the network was present at the last scattering surface (LSS). For simplicity, we consider a $Z_2$ model, with two degenerate vacua. As is well known, a DW network has an attractor `scaling' regime where at all times there is about one Hubble-sized DW per Hubble patch. The LSS, then, is divided into many Hubble patches, with about one DW in each. To visualize the relevant information (the field excursion along each line of sight), it is useful to draw a ``celestial plot" showing the field as a function of the angle and redshift.
For illustration, we sketch in Fig. \ref{fig:sketchZ2} how such a celestial plot must look like in a model with two vacua in scaling. 
In scaling, the DW network shapes are basically random while the volume in different vacua remains balanced. Then, the CMB photons that we observe across the sky would come from regions of the different vacua with equal probabilities.
However, our local universe spontaneously breaks the symmetry and selects one of the minima. Then, the sky-average of $\Delta\phi$ along the line of sight is nonzero and, as it turns out, it gives the right order of magnitude for the CMB isotropic birefringence  \cite{Takahashi:2020tqv}. 
Fig. \ref{fig:sketchZ2} can be readily compared with Fig. \ref{fig: past light cone} where we show the same quantity but using instead real data obtained from of one of our 3D numerical simulations in a $Z_2$ model with degenerate vacua.
Despite that scaling is only reached after some time, the plot with real data confirms that at all redshift $z$, there are statistically equal numbers of Hubble patches of either color.

However, as we argue next, the nonzero isotropic birefringence signal can actually be achieved in any axionic defect network, both with DWs and/or strings (cf. Fig. \ref{fig:sketch-strings}).
Concerning the axionic string and string-wall networks, recent works have discussed the isotropic birefringence  and concluded that the isotropic signal cannot be obtained in this case because it would violate current upper bounds on the anisotropic component \cite{Agrawal:2019lkr,Jain:2021shf,Jain:2022jrp}. 
These works have focused on the birefringence produced by the spatial and time distribution of string loops. However, here we will propose three ways that can potentially circumvent this no-go result.
First, the presence of a defect, either a string loop or a closed domain wall, not too far from us so that it encompasses a sizeable subtended angle in the sky. Second, the presence of gradients of the axion field in each Hubble patch that provides an additional field excursion; for axionic strings the axion field is massless and so large gradients in any Hubble patch are unavoidable.
Third, if the string network only forms and or annihilates between recombination and today the anisotropic signal can be suppressed. 
Note that, in the first two options, the signal is of environmental type, in the sense that observers located at different places in our present Hubble patch would find different values.

Regarding the anisotropic component of the signal, there is
no evidence yet; so far only upper bounds have been derived \cite{2012PhRvDWMAP7,POLARBEAR:2015ktq,BICEP2:2017lpa,Namikawa:2020ffr,SPT:2020cxx,Gruppuso:2020kfy,Bortolami:2022whx},  however, the situation might change with future data and the increase in sensitivity \cite{Pogosian:2019jbt}.
Different scenarios predict different levels of anisotropic birefringence; while in the ultra-light axion case, the anisotropies in the signal will likely be small, if of inflationary origin \cite{Caldwell:2011pu,Fujita:2020aqt}, in the case of axionic networks the anisotropies are expected to leave a sharp footprint in the spectrum \cite{Agrawal:2019lkr,Kitajima:2022jzz,Gonzalez:2022mcx}.
We will compute those anisotropies by taking as a proxy dedicated 3D numerical simulations of DW networks in a $Z_2$ model with degenerate vacua and highlight the crucial role of the reionization component of the spectrum. Moreover, the anisotropic birefringence spectrum, similarly to the isotropic component, carries information about the time dependence of the signal. There is also no evidence for the time dependence but with future CMB experiments such as Simons Observatory \cite{SimonsObservatory:2018koc}, LiteBIRD \cite{Matsumura:2013aja,LiteBIRD:2022cnt} or CMB-S4 \cite{CMB-S4:2016ple,CMBS4} it will be possible to perform a tomographic approach by probing the spectral shape of the $EB$ correlation function with different angular scales and thereby measuring the amounts of birefringence angle at different times, the recombination and reionization epoques \cite{Sherwin:2021vgb,Nakatsuka:2022epj,Lee:2022udm,Galaverni:2023zhv,Namikawa:2023zux}.
We will discuss how the tomographic approach could allow to distinguish the formation/annihilation times of the network.

The discussion so far has been largely independent of the  tension of the walls in the DW and string-wall networks. However, if the tension of the walls is large enough the network will also leave a gravitational imprint on the CMB.
In the  scaling regime,   
the network redshifts slower than matter and radiation and therefore has a tendency to dominate the universe, what is commonly known as the \textit{DW problem} \cite{Zeldovich:1974uw}. This problem can however be avoided at least in two ways: i) the tension of the DW is so small that the network is irrelevant at all times in the cosmic history; ii) the network annihilates before it dominates the universe. 
The annihilation of the network is not an exotic possibility. Anomalous contributions from non-perturbative sectors (see e.g. \cite{Sikivie:1982qv,Kamionkowski:1992mf}) or even the standard lore that global symmetries are broken by quantum gravity motivate a soft breaking of the discrete symmetry that would cause pressure on the walls and annihilate the network; and several other mechanisms have also been proposed \cite{Vilenkin:1981zs,Coulson:1995nv,Babichev:2021uvl,Gonzalez:2022mcx}.

The fact that the network annihilates before the present time does not hinder the possibility of probing it. 
DW and string-wall networks in the scaling regime become energetically more and more relevant as time passes by, and  so if long-lived they can generate a loud spectrum of stochastic gravitational waves (GWs). Recent studies have shown that the GW signal generated by the network can be probed by current \cite{ZambujalFerreira:2021cte} and future GW observatories (see \cite{Saikawa:2017hiv} for a review), and potentially even explain the spectrum observed by pulsar timing array observatories \cite{Ferreira:2022zzo,Kitajima:2023cek}.
However, the regime where the network annihilates after recombination and the GW spectrum peaks at CMB scales has remained so far unexplored. In this work, we will compute the GW spectrum from dedicated numerical simulations and, using previous CMB searches for stochastic gravitational wave backgrounds \cite{Namikawa:2019tax}, derive the first CMB bounds on  networks with DWs that annihilate after recombination thus closing the existing gap in the literature: previous constraints applied to networks that either annihilate before recombination \cite{Gelmini:2021yzu,Ferreira:2022zzo,Ramberg:2022irf} or survive until today \cite{Zeldovich:1974uw,Lazanu:2015fua,Sousa:2015cqa}. We will finish this work by discussing how future measurements of CMB B modes will improve the sensitivity to these networks.

The paper is organized as follows. In Section \ref{sec: DW networks}, we describe the basic properties of a DW network and discuss different types of topological defect networks that can be hosted in the axionic case.  We then discuss in Section \ref{sec:CMB Isotropic birefringence}   the  CMB isotropic birefringence signal that is characteristic of the different networks and in Section \ref{sec:Anisotropic birefringence} we compute the associated anisotropic birefringent signal by taking as a proxy the results from our DW network simulation.  In Section \ref{sec:Tomography} we extend the discussion to the case of networks that form or annihilate between the LSS and today. Finally, in Section \ref{sec: GWs} we study the GWs generated by the network and derive bounds on the DW parameters from the non-observation of CMB B-modes and conclude in Section \ref{sec: Conclusion}.

\section{Basics of defect networks}
\label{sec: DW networks}

In this section, we start by briefly recapping the basic properties of domain wall networks in the scaling regime and then discuss the type of defect networks that can be found in axionic case.

\subsection{Domain wall networks}
The spontaneous symmetry breaking of discrete symmetries in the early universe is known to lead to the formation of DW networks \cite{Kibble:1976sj}. 
DWs are the localized field configurations that bridge the field profile between  (quasi-)degenerated minima. They are characterized by a width $L$, related to the mass $m_\phi$ of the field by $L\sim 1/m_\phi$, and by a tension $\sigma=\int  \phi'(z)^2 dz $ where $z$ is the direction orthogonal to the wall.

Roughly speaking, the network forms when the wall width fits in one Hubble radius, $ m_\phi(z_f) \simeq H(z_f)$ where $z_f$ is the redshift of formation. From that point on, the continuous dissipation of substructure leads the network to an attractor regime, known as scaling, that features some universal properties (like the scalar and GW spectrum that we will explore in the next sections) that are expected to be mostly insensitive to the initial conditions.
While in scaling, the network has energy density  (see e.g. \cite{Vilenkin:2000jqa})
\begin{equation}
	\rho_\text{dw}= c \, \sigma H \label{eq:rho_dw}
\end{equation}
with $c \sim {\cal O}(1)$ the average number of DWs per Hubble patch that here we take from the numerical simulations described in App. \ref{app: Numerical Simulation}.
The DW energy density dilutes slower than matter or radiation thus causing the abundance 
\begin{eqnarray} \label{eq:DW abundance}
	\Omega_\text{dw}(\eta)\equiv \frac{\rho_\text{dw}}{\rho_\text{tot}} = \frac{c \, \sigma}{3M_p^2 H}
\end{eqnarray}
to increase over time, where $\eta$ is the conformal time coordinate.
This remark will be important in Section \ref{sec: GWs} when we study signatures that are proportional to $\Omega_\text{dw}$, hence, maximal around the latest time the network is in scaling. 

A long-lived network is therefore a dangerous cosmological relic. However, even if long-lived, the network can still annihilate before it dominates the energy budget of the universe. That can happen for example if: i) there is a restoration of the discrete symmetry at low energies causing the field to be massless and the walls to disappear \cite{Vilenkin:1981zs,Babichev:2021uvl}; ii) there is a bias in the initial conditions that populates one minimum more than others \cite{Coulson:1995nv,Larsson:1996sp,Hindmarsh:1996xv,Gonzalez:2022mcx}; iii) the discrete symmetry is only approximate and the differences in vacuum energy between the different minima exert pressure on the walls and make them collapse \cite{Sikivie:1982qv}.

In this work, we will keep the analysis agnostic with respect to the annihilation mechanism and simply define the time (redshift) of annihilation of the network as $\eta_\text{ann}\, (z_\text{ann})$, or equivalently the corresponding CMB temperature $T_\text{ann}$. This time scale can then be related to the physical parameters causing the annihilation of the network in the different mechanisms.
Furthermore, we will work in a sudden decay approximation where we assume that the network annihilates quickly and neglect additional signals that can be generated  during the annihilation process, where the scaling properties are no longer verified. In this sense, our approach will be conservative as the annihilation stage can further enhance the signals but not suppress them.

\subsection{Axionic defects}

The distinctive feature of the axionic case is that essentially the axion is an angular variable. This immediately leads to the possibility of having axionic cosmic strings. Moreover, an axion potential is allowed, provided it is periodic. To fix ideas, we can take it to be
\begin{equation}\label{pot}
V(\phi) = \frac{f^2 m^2}{N^2} \left[1-\cos\left(N \phi/f\right)\right]~  
\end{equation}
with integer $N$. One recognises at once i) a discretum of degenerate minima $a=2\pi i f/N$ with integer $i$;  ii) the possible existence of DWs, interpolating between neighboring vacua.

Thus, a given axion low energy theory can host a veriety of topological defect networks. Depending on when the corresponding Peccei-Quinn (PQ) symmetry (that gives rise to the axion) is broken, one can envisage 4 types of network, consisting of: 
\begin{itemize}
    \item[$\star$] {\it Strings only}. Strings arise in post-inflationary models, when the PQ symmetry is spontaneously broken after inflation. Axionic string networks seem to reach a scaling regime, roughly characterized by a $O(1)$ long string per Hubble patch. 
    Our discussion below is not very sensitive to possible logarithmic deviations from scaling.

    \item[$\star$] {\it Strings and DWs}. The `explicit' PQ breaking term in Eq. \eqref{pot} leads to $N$ walls attached to each string. Previous numerical simulations \cite{Hiramatsu:2010yz,Kawasaki:2014sqa} show that this type of network reaches scaling in the absence of a large `bias' (further explicit breaking) terms. To a good enough approximation, there is around about one string and $N$ walls per Hubble patch implying that all $N$ vacua are statistically equally represented.  

    \item[$\star$] {\it DWs only}. A network of axionic DWs (without strings) can be realized if, for example, the $U(1)$ symmetry giving rise to the axion is broken during inflation and there is enough quantum diffusion, proportional to $H_\text{inf}/f$ where $H_\text{inf}$ is the Hubble rate during inflation, such that the field excursion throughout inflation is larger than one period ($2\pi f$) and the network forms at late times \cite{Takahashi:2020tqv,Kitajima:2022jzz,Gonzalez:2022mcx}. 
    
    \item[$\star$] {\it Annihilating networks}. Any additional source of the PQ symmetry leads to a potential of the form \eqref{pot}, but generically aligned at a different field value and with a different value of $N$. This breaks the degeneracy between vacua, causes the annihilation of the network and opens up the possibility that the network was present at the last scattering surface (LSS) but disappears at some point in between us and the LSS.
 
\end{itemize}
In the next section, we discuss how the isotropic birefringence signal arises in all these cases. In brief, for networks with DWs a picture like Fig.~\ref{fig:sketchZ2} applies whereas networks with strings are better described by the sketch in Fig. \ref{fig:sketch-strings}.

\section{Isotropic birefringence}
\label{sec:CMB birefringence}

We now move to the phenomenon of birefringence. We start by discussing the isotropic component and leave the discussion the anisotropies to Sec. \ref{sec:Anisotropic birefringence}.

Let us assume that the field $\phi$ couples to photons through the topological term,
\begin{equation}\label{eq: axial coupling}
	c_\gamma \frac{\alpha_\text{em}}{8\pi}  \frac{\phi}{f}  F_{\mu \nu}\tilde{F}^{\mu \nu}
\end{equation}
where $\tilde{F}^{\mu \nu}=\epsilon^{\alpha \beta \mu \nu} F_{\alpha \beta}/(2\sqrt{-g})$, with $\epsilon$ is the Levi-Civita tensor, and $g$ the determinant of the metric. The dimensionful coupling $f$ parametrizes the strength of the interaction, $\alpha_{\rm em} \simeq 1/137$ is the fine structure constant and $c_\gamma$ is a dimensionless coefficient. In the case of an axion-like particle, $c_\gamma$ is typically  $E/N$ where $E$ is the electromagnetic anomaly and $N$  the DW number.

The coupling in eq. \ref{eq: axial coupling} acts as a birefringent medium  for the CMB photons thus causing the rotation of the polarization axis along their path.
The amount of rotation $\beta$ that is generated from a time $\eta$ (redshift $z$) until the present time $\eta_0$ ($z=0$) is \cite{Carroll:1989vb}
\begin{equation} \label{eq: Delta theta general}
\beta(z)=\int_{\eta_0}^{\eta(z)}c_{\gamma}\frac{\alpha_{\rm em}}{4\pi} \Dot{\left<\theta\right>} \,d\eta=\frac{\alpha_{\rm em}c_{\gamma}}{4\pi} \left(\langle\theta(z)\rangle-\theta_\text{loc} \right) \approx 0.21 c_\gamma \parfrac{\langle\theta(z)\rangle-\theta_\text{loc}}{2\pi}\, \text{deg}
\end{equation}
where $\theta \equiv \phi/f$ and  $\left< \theta(z) \right>$ denotes the average of $\theta$ over the observable universe at redshift $z$
and $\theta_\text{loc}$ is our local value. 
The CMB data is sensitive to the rotation that takes place after recombination  mostly through the parity-odd cross-correlation power spectrum
 $C_\ell^{EB} \propto \sin(4\beta)C_\ell^{EE} $ that would vanish in the absence of birefringence \cite{Lue:1998mq}. 

We proceed by first describing the 
expected values for $\beta_\text{rec}\equiv \beta(z=z_\text{rec})$ from networks in the scaling regime, first for a $Z_2$ DW network and then for the axionic defect networks with strings assuming they are present at recombination and are stable until today. The case of annihilating and late-forming networks is discussed in Sec. \ref{sec:Tomography}.

\subsection{$Z_2$ Domain Walls}
\label{sec:CMB Isotropic birefringence}

A $Z_2$ domain wall network (without strings) in the scaling regime at recombination, $m_\phi \gtrsim H_\text{rec} =3 \times 10^{-29}$eV, is characterized by about one wall per Hubble patch.
Therefore, at recombination,
the field is expected to equally populate the $N=2$ minima at $\theta_i= 2\pi i/N$,
and the probability that a photon coming from a solid angle $\Omega$ originates from a region where $\theta$ is at the minimum $i$ is then $p(\theta_i,\Omega)=1/N$
(see Figs. \ref{fig:sketchZ2} and \ref{fig: past light cone} for a visual confirmation of this equal-sharing property).

With these considerations in mind, the amount of birefringent rotation from recombination is  
\begin{eqnarray}
 \beta_\text{rec}  &=& \frac{\alpha_\text{em} c_\gamma}{4\pi}  \left(\int \frac{d\Omega}{4\pi}\, \sum_{i=1}^N  \theta_i \,p(\theta_i,\Omega) - \theta_\text{loc} \right)=   \frac{\alpha_\text{em} c_\gamma}{4\pi} \left(\sum_{i=1}^N  \frac{2\pi i}{N} - \theta_\text{loc} \right) \\ &=& \frac{\alpha_\text{em} c_\gamma}{2} \left[\frac{N+1}{N} - \frac{\theta_\text{loc}}{2\pi} \right] \, .
\end{eqnarray}
The prediction for $\beta_\text{rec}$ is probabilistic; it is equally probable for our local universe to be at any of the minima, $p(\theta_{i_0}=\theta_i)=1/N$. 
For $N=2$, there are two possible outcomes for isotropic birefringence, $ \beta_\text{rec}\simeq \pm 0.21 c_\gamma/2$  degrees,  each with 50\% probability \cite{Takahashi:2020tqv,Kitajima:2022jzz}.  

\subsection{String and string-wall networks}
\label{sec: The role of strings}

When strings are present the signal is richer and can be split into two types of contributions:   from the crossing of defects such as string loops $\beta_\text{loops}$ and from gradients $\beta_\text{grad}$.
The first type has been considered in previous works \cite{Agrawal:2019lkr,Jain:2021shf,Jain:2022jrp} and it gives a $\pm2\pi$ shift in the axion field excursion every time the photon crosses an axion string loop.  The gradient component
also contributes to the field excursion between the LSS $ \theta_\text{LSS}$ and the value today $\theta_\text{loc}$ and we can actually define it modulo $2\pi$ to remove the effect of the loops.
Therefore, putting both contributions together we have
\begin{eqnarray}
    \beta = \beta_{\text{grad, mod} \, 2\pi} + \beta_\text{loops} \, .
    \end{eqnarray} 
Each contribution moves the axion field, and consequently rotates the photon polarization, randomly in both positive and negative directions thus we can think of the birefringence along each line of sight as being a realization of this random process\footnote{In saying this we have implicitly assumed that the mean of the distribution of the axion field at the LSS corresponds to the value in our local universe. If this assumption does not hold there would be a further contribution to the isotropic signal coming from $\left<\theta \right>_\text{LSS} - \theta_\text{loc}$.} and, as we explain next in more detail, the lowest multipoles are those contributing to the isotropic birefringence thus causing an effect of the ``environmental" type.

Let us see that more explicitly in the case of string loops. In scaling, we expect around one loop in each Hubble patch at all times so photons, coming from each line of sight $\hat{n}$, can pass through several randomly oriented loops. The number of crossed loops is about the number of e-folds to LSS, $N_e(z_{\rm LSS})$. Therefore, in each direction, the polarization of the CMB experiences a random walk of $N_e(z_{\rm LSS})$ steps of  $\pm 2\pi$ in $\Delta \theta$ which gives a variance of  $\langle\Delta\theta(\hat{n})^2\rangle^{1/2}\simeq 2\pi\sqrt{N_{e}(z_{\rm{LSS}})}$.
However, to compute the monopole we need to average over all angles. At the LSS, the total number of independent realizations of this random walk is roughly given by the number of Hubble patches at the LSS\footnote{The number of patches is computed from the angle subtended by the horizon at the LSS $\eta_{\rm LSS}/\eta_0\simeq0.022$ rad, thus $N_{\rm{patches}}\simeq (2\eta_0/\eta_{\rm LSS})^2$. }, which is $N_{\rm{patches}}\simeq 7\times10^3$. Thus, the variance of the distribution for 
 $\Delta \theta$ can be estimated from the central limit theorem, to be $\langle\Delta\theta_{\rm{LSS}}^2\rangle ^{1/2} \simeq 2\pi\sqrt{N_{e}(z_{\rm{LSS}})/N_{\rm{patches}}(z_{\rm{LSS}})}\ll1$.
This reasoning can be repeated for different opening angles and redshifts by accounting for the fact that $N_{e}(z)$ is the number of e-folds up to redshift $z$ and the number of patches at that time depends on the angle subtended by the horizon at redshift $z$. Then it becomes clear that the major contribution to the variance $\langle\Delta\theta_{\rm LSS}^2\rangle$ comes from low redshifts/ large opening angles where $\sqrt{N_{e}/N_{\rm{patches}}}\sim \mathcal{O}(0.1-1)$ and can be understood as the contribution from the nearest structures in our local universe thus causing an effect of the ``environmental" type. Since the closest structures contribute the most to the isotropic birefringence, we should also take into account other types of ``environmental" effects that might be subdominant for computing the anisotropic power spectrum on large scales.

\begin{figure}
    \centering
\includegraphics[width=0.44\linewidth]{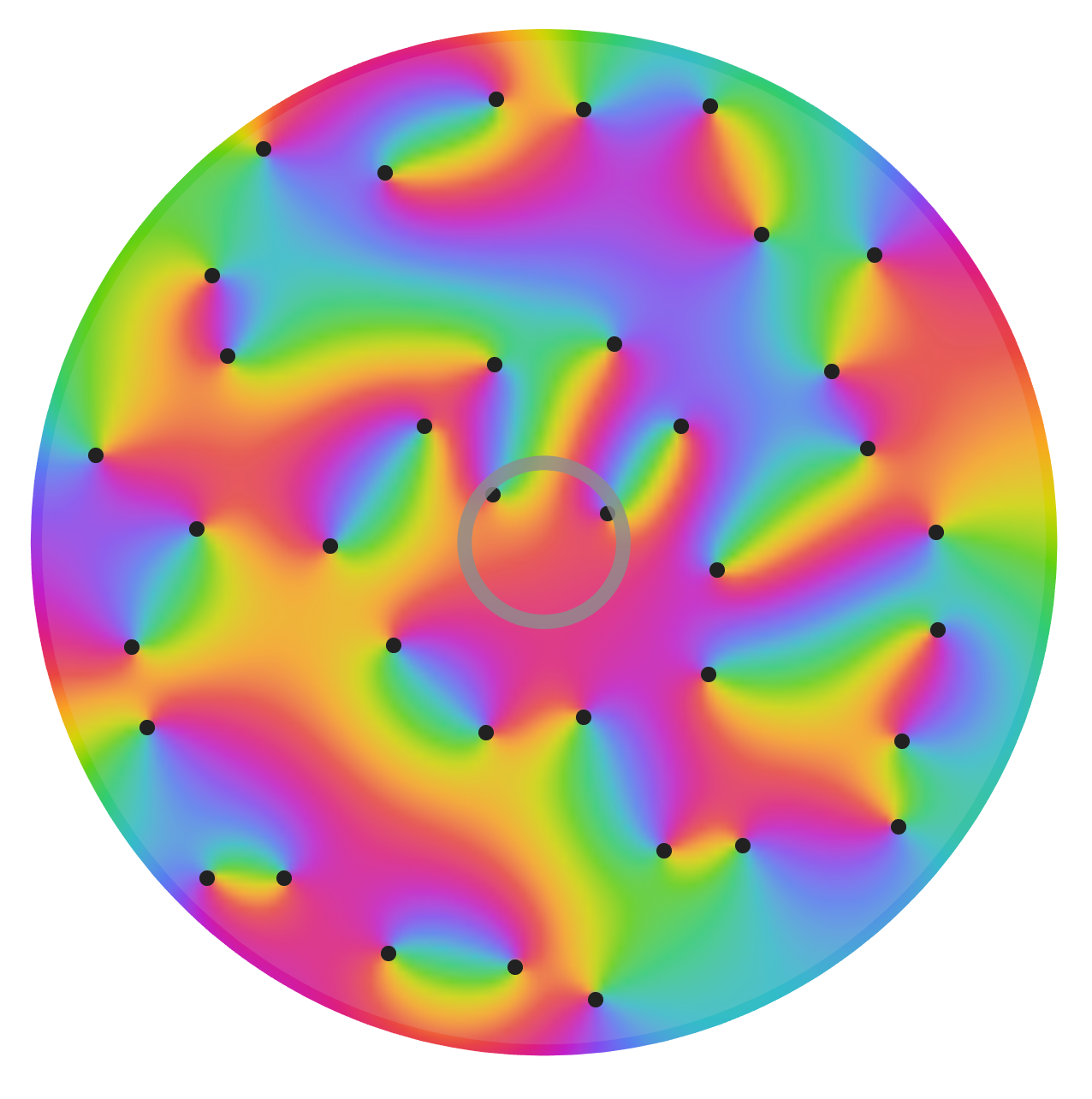}
\includegraphics[width=0.47\linewidth]{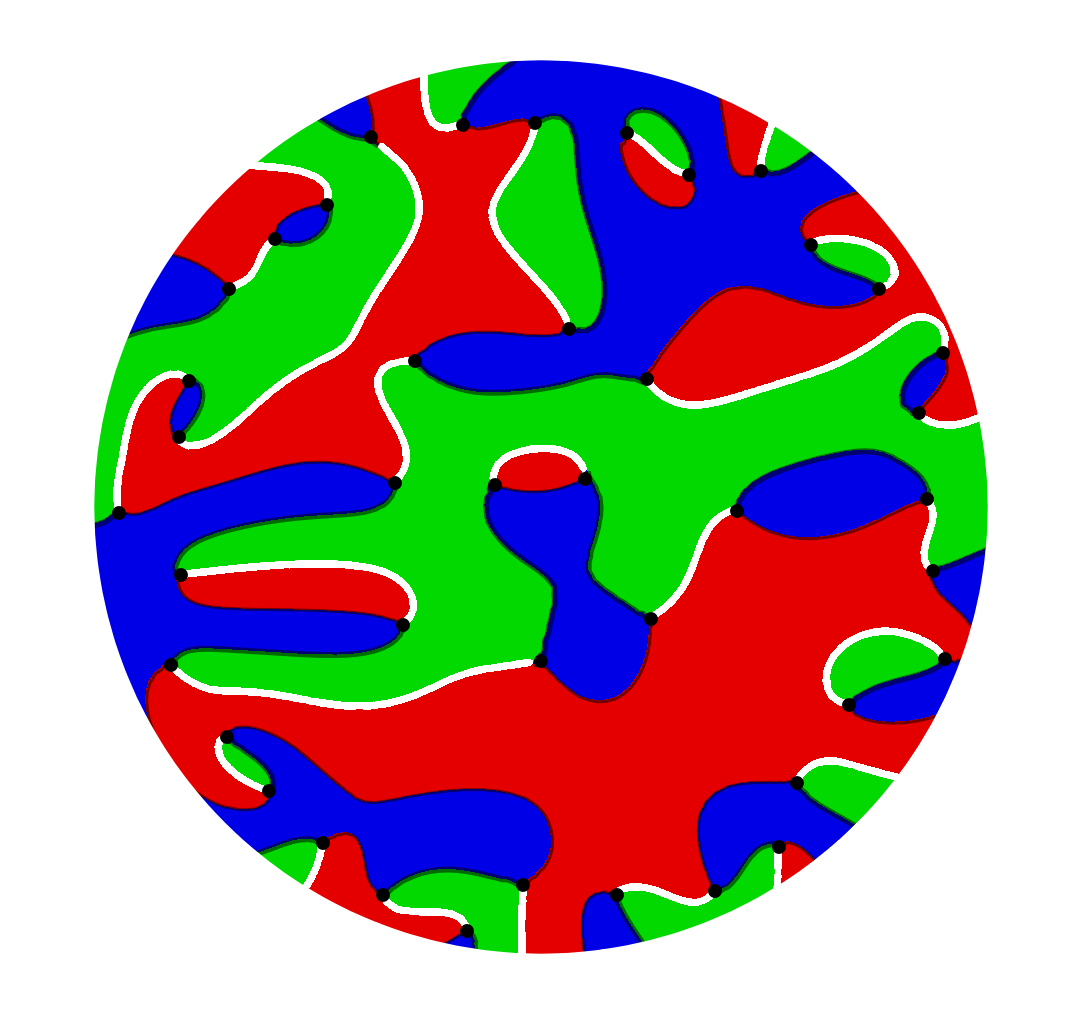}
    \caption{Illustration of the environmental effect in a cosmic string network with DWs (right) and without (left). In a 2D cut, the strings can be represented as `vortices'. Colour corresponds to the value of the axion field. Colour flows in red-green-blue order (counter-)clockwise around (anti-) vortices. As in the previous figures, the radial distance corresponds to redshift, earlier times at the edge of the disk -- which represents the LSS. The inner gray circle represents a present-day Hubble patch (say, $z=1$). The precise location of, say, the solar system is a random choice in the gray circle that is completely uncorrelated from the axion configuration at LSS. This leads to a uniform distribution of values and an $O(\pi)$ estimate for its variance. The case of a string-wall network with domain wall number $N=3$ is shown on the right. }
    \label{fig:sketch-strings} 
\end{figure}

For example, let us consider the effect of a single nearby loop.
 The axion field receives a $\pm 2\pi$ boost in the solid angle $\Omega$ enclosed by the loop.
Then we have that
\begin{eqnarray}
   \beta_\text{loop} = 0.21 c_\gamma \left(\frac{\Omega}{4\pi}\right) \text{ deg.}
    \end{eqnarray} 
The coefficient $c_\gamma$  is bounded by the anisotropic birefringent effect that string loops have on the CMB \cite{Agrawal:2019lkr,Jain:2021shf,Jain:2022jrp}. For a network with  one Hubble-size string loops per Hubble patch, the axion-photon coupling is constrained to be $c_\gamma \lesssim 3.8$ \cite{Jain:2022jrp}. Therefore, a nearby string loop could explain the isotropic birefringent signal at the $2\sigma$ level if it encloses an angle of at least $2.5$ steradians, around $1/5$ of the whole sky.
Clearly, the likelihood that a loop encloses this solid angle clearly depends on the size of the loop. For small loops this would certainly require a quite special location. However, a network that survives until today typically contains Hubble-sized loops that can cover a significant solid angle even at a cosmological distance from us.
For example, in order to cover 2.5 steradians, a Hubble-sized loop should be no farther than redshift $z\sim 1/\tan(2.5/2)\sim 0.3$.
This seems compatible with the configurations encountered during scaling, typical distance between strings (and large loops) being of order $1/H$.
To better quantify this statement it would be interesting to study the probability distribution for a random point to be at a given distance from the closest loop.

In \cite{Jain:2022jrp} the authors found that when accounting for the cumulative effect of loops at higher and higher  redshifts, the typical size of the isotropic signal was not enough to explain the value recently observed (without spoiling the constraints on the anisotropic signal). It would be interesting to see if the inclusion of closer structures (lower redshifts) in the analysis could change the conclusions or if the cancellation of the loop effect from larger redshifts could be reduced when allowing for a distribution of loop sizes (the authors assumed a fixed comoving loop size at all redshifts).

Yet, just an order two increase in the variance of the isotropic signal from string loops found in \cite{Jain:2022jrp}  would be enough to avoid the no-go conclusion. Therefore, it is important to understand the role played by the remaining contributions, the gradients, to isotropic birefringence.
We identify two kinds of gradients depending on whether the network has DWs or not. 
In the case with DWs, our local universe could, for example, be enclosed by a wall that is attached to a string loop (as in Fig. 3 of \cite{Jain:2022jrp}).
Photons that come from the outside of the domain wall, coherently feel an axion field excursion proportional to $2\pi/N$, because they had to cross a few walls, whereas those that come from the direction of the string loop feel instead an excursion of $-2\pi+2\pi/N$. Assuming that the string loop encloses a much smaller angle in the sky than that of the enclosed DW then the typical size of the signal would be
\begin{eqnarray}
    \beta_\text{enclosing DWs} \simeq 0.21 c_\gamma/N 
\end{eqnarray}
A similar result can be found if we think instead of a domain wall that cuts asymmetrically our current universe.

In the absence of DWs, the axion field is massless and so it can also develop large gradients even more so if there is a defect not too far from our local universe.  
For example, as also noted in \cite{Agrawal:2019lkr}, if a photon moves on a trajectory perpendicular to a straight string it will travel through an axion field excursion of $\pi$ (asymptotically). This field excursion needs not to be exactly dipolar over the whole sky, which would be the case if, in each half of the sky, there is an $\pm$ excursion 
 as there can naturally be some asymmetry. Therefore, we estimate the environmental contribution of gradients to the monopole to be
\begin{eqnarray}
    \beta_{\text{gradients, mod} \, 2\pi} \simeq 0.21 c_\gamma/2 \times \nu 
\end{eqnarray}
where $\nu$ is the portion of the sky with the asymmetry (see Fig. \ref{fig:sketch-strings} for an idea of the environmental effects). 

These additional contributions to isotropic birefringence, which can be of the same order of magnitude as those for string loops, will unavoidably increase the signal, potentially, without spoiling the anisotropic measurement. It is however crucial to perform a more dedicated analysis of this situation, optimally using lattice simulations of cosmic strings, to confirm if that is indeed the case. A smoking gun for these ``environmental" effects from stable networks would be that the isotropic birefringence turns to be driven by data coming from a particular direction of the sky.

On the other hand, possible logarithmic deviations from scaling would, if confirmed, increase the average number of strings per Hubble patch by a number $\xi \sim {\cal O } (\log(f/H))$, and potentially affect the results in two ways. First, the anisotropic constraint on $c_\gamma$ would become stronger by a factor of $1/\sqrt{\xi}$ thus reducing the maximal $\beta$ allowed. Second, because there would be more strings at low redshifts their contributions to the isotropic birefringent signal would partially cancel and suppress the signal by a factor of $1/\sqrt{\xi}$. These two effects combined can therefore suppress the maximal $\beta$ by a factor of $1/\xi$. However, apart from the uncertainty on the logarithmic departure from scaling, and whether it can be extrapolated for a large hierarchy of scales, the effect is also model-dependent because of the dependence on $f$. For sufficiently small $f$ the logarithmic effect even if present will give a negligible contribution.

In Sec. \ref{sec:Tomography}, we extend the discussion to the case where the network forms or annihilates in between recombination and today but we anticipate that also in that case there can still be a contribution to the isotropic birefringent angle from the misaligned contributions to the potential that cause the network to annihilate.

\section{Anisotropic birefringence}
\label{sec:Anisotropic birefringence}

The spatial distribution of the defects inevitably generates an angular-dependent signal $ \beta(\eta,\mathbf{\Hat{n}})$  on top of the isotropic component given in eq. \eqref{eq: Delta theta general} which is expected to be much larger than in the pre-inflationary case where anisotropies are suppressed by $H_I/f_a\ll1$ with $H_I$ the Hubble scale during inflation \cite{Kolb:1990vq}. Therefore, looking for anisotropies in the birefringence is better motivated in the presence of topological defects. This anisotropic component induces mixings between the $C^{EE},C^{BB}$ and $C^{EB}$ spectra, similarly to the effect of isotropic birefringence, but involving different multipoles \cite{Kamionkowski:2008fp,2009PhRvD..80b3510G,Gubitosi:2011ue}.
Contrary to the isotropic birefringence, no evidence of anisotropic birefringence has been measured and current BICEP/Keck data puts a tightest bound on its magnitude $\ell(\ell+1)C^{\beta\beta}_\ell/2\pi\leq 0.014$ $\text{deg}^2$ \cite{BICEPKeck:2022kci}, improving on previous results from the Atacama telescope and SPT mission $\leq 0.033$ $\text{deg}^2$ \cite{Namikawa:2020ffr,SPT:2020cxx}. However, these bounds assume a scale-invariant power spectrum; for defects, one should compare the data with the expected $C^{\beta\beta}_\ell$ spectrum for which we will provide a good fit in the Appendix \ref{app:FittingCl} that will  hopefully be useful for future searches.

In what follows, we review the formalism that connects the anisotropic birefringence spectrum with the scalar power spectrum. We then take the power spectrum from 3D numerical simulations of $Z_2$ DW networks, whose details are discussed in \ref{app: Numerical Simulation}. We interpret these results as a proxy for the axionic defects. Networks with strings naturally differ from the DW-only case in several aspects, like in the presence of string loops. However, the presence of a peak in the spectrum at horizon scales is also expected in the cases with strings therefore justifying the use of the DW network as a proxy.

\subsection{Scalar power spectrum}

\label{sec: Scalar power spectrum}
\begin{figure}
    \centering
    \includegraphics[scale=0.7]{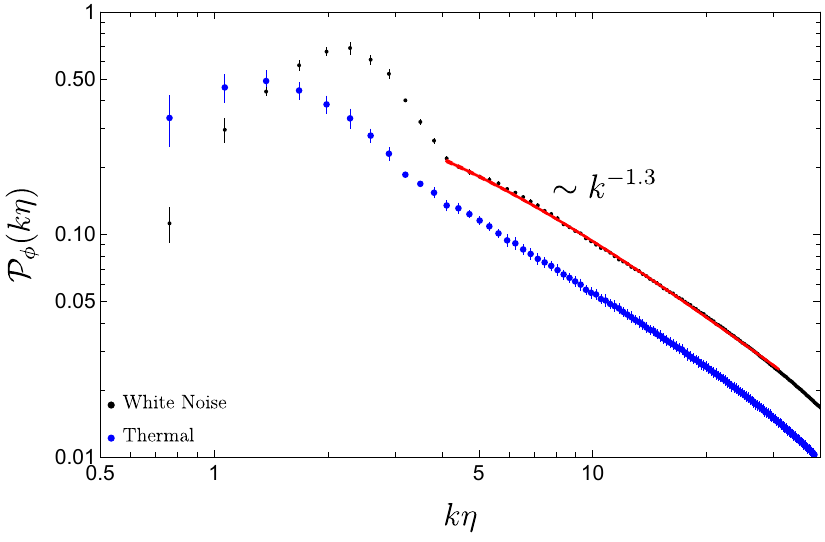}
    \caption{3D scalar power spectrum of the $Z_2$ DW network. In blue we show $\mathcal{P}_\phi$ for the thermal initial condition and in black that for the white noise initial condition. Note that on scales smaller than the peak the spectrum has the same scaling whereas on larger scales depends on the choice of the initial conditions. }
    \label{fig:powerspec}
\end{figure}

The first quantity we should look at is the scalar power spectrum for the DW network $P^\theta_{k_1}(\eta)$, defined as  
\begin{equation} \label{eq: Scalar field power spectrum}
\langle \theta^*(\eta,\mathbf{\hat{k}}_1)\theta(\eta,\mathbf{\hat{k}}_2)\rangle=(2\pi)^3\delta^3(\mathbf{k}_1-\mathbf{k}_2) P^\theta_{k_1}(\eta) 
\end{equation}
and related to  the adimensional quantity $\mathcal{P}^\theta_k=k^3P^\theta_k/(2\pi^2)$.
In the scaling regime, the spectrum only depends on $k\eta$ and presents a constant peak at horizon scales $k\eta\sim1$ as shown in Figure \ref{fig:powerspec} for two different common choices of initial conditions, white noise (black) and thermal (blue) described in \ref{app: Numerical Simulation}. The spectrum has a universal behaviour $\sim k^{-1.3} $ at scales smaller than the horizon that is close, but slightly deviates from the analytical expectation $\sim k^{-1}$  given in \cite{Kitajima:2022jzz,Takahashi:2020tqv} where the 2D network is approximated by a random process following a Poisson distribution in real space. Note that in the thermal case the amplitude at small scales is suppressed by an order $\mathcal{O}(1)$ compared to the white noise case. On the other hand, at scales larger than the horizon the slope of the power spectrum depends on the initial condition with the thermal case being more similar to the inflationary initial conditions  \cite{Gonzalez:2022mcx}, due to the assumption of larger power on superhorizon scales in the initial conditions.
In what follows we focus on the anisotropic power spectrum for the white noise case, but  stress that the results on large scales/small multipoles can vary if the initial conditions feature correlations on superhorizon scales (see also \cite{Gonzalez:2022mcx}). 

\subsection{Anisotropic birefringence from recombination and reionization}

To connect the anisotropic birefringence to the scalar power spectrum, we start by decomposing in spherical harmonics the birefringent angle at a time $\eta$ for a photon coming from the direction $\hat{n}$ as
\begin{equation}\label{eq:betadeco}
\beta(\eta,\mathbf{\Hat{n}})\equiv \left(\frac{c_\gamma\alpha_\text{em}}{4\pi }\right)\Delta\theta(\eta,\mathbf{\Hat{n}})= \sum_{\ell,m}\beta_{\ell m}(\eta)Y_{\ell m}(\mathbf{\Hat{n}})
\end{equation}  
where  $\Delta\theta(\eta,\mathbf{\Hat{n}})=\theta(\mathbf{\Hat{n}},\eta)-\theta(\eta_0)$. 
The anisotropic birefringence power spectrum $C^{\beta\beta}_\ell $ is then given by
\begin{equation}
    \langle \beta^*_{\ell_1 m_1}(\eta_1)\beta_{\ell_2 m_2}(\eta_2)\rangle=C^{\beta\beta}_{\ell_1} (\eta_1, \eta_2) \delta_{\ell_1\ell_2}\delta_{m_1 m_2}, 
\end{equation}
which, as we review in Appendix \ref{app:angular PS}, is related to the scalar field power spectrum $P^\theta_{k}(\eta)$ via \cite{Greco:2022xwj} 
\begin{equation}\label{eq:Cbeta}
    C^{\beta\beta}_\ell  (\eta_1, \eta_2)= 4\pi \left(\frac{c_\gamma\alpha_\text{em}}{4\pi }\right)^2 \int\frac{{\rm d}k}{2\pi^2}k^2j_{\ell}(k\Delta\eta_1)j_{\ell}(k\Delta\eta_2)P_{k}^\theta(\eta_1,\eta_2),
\end{equation}
where $\Delta \eta_i\equiv \eta_0-\eta_i  $ is the comoving distance. Note that by taking $\beta\simeq c_\gamma \alpha_{\text{em}}/4$, valid for the $\mathcal{Z}_2$ model we are considering, the anisotropic and the isotropic signal are correlated $C^{\beta\beta}_\ell\propto\beta^2$.

Since the polarization of the CMB photons is mostly generated at two instances of time, recombination and, to a lesser extent, reionization, the anisotropic birefringence spectrum will mainly have contributions from these two epochs. 
Following \cite{Sherwin:2021vgb,Greco:2022xwj}, we treat both rotations as independent and decompose the birefringence angle as\footnote{The recombination and reionization contributions can be isolated by convoluting the power spectrum with the visibility function around the recombination and reionization peaks \cite{Greco:2022xwj}. In this work, for simplicity, we work under the instantaneous emission approximation, in this way, we can provide some useful formulas that can be used to compare with CMB data, as through eq. (67) of \cite{Greco:2022xwj}. As we discuss below in more detail, we expect this simplification to have negligible effects on the recombination piece, due to the sharpness of the visibility function at that time, but perhaps a non-negligible effect on the reionization contribution. Moreover, we note that taking into account the full redshift evolution of the anisotropies can leave non-trivial effects. Indeed, the anisotropies grow until the network reaches scaling, thus, a bigger amplitude at later times can compensate for the smallness of the visibility function outside the recombination and reionization period. However, here we do not consider such a possibility.}:
\begin{equation}
\beta(\mathbf{\Hat{n}})=\sum_{\ell,m} (\beta^{\text{rec}}_{\ell m}+\beta^{\text{rei}}_{\ell m}) Y_{\ell m}(\mathbf{\Hat{n}})
\end{equation}
so the total power spectrum presents the following contributions
\begin{align} \label{eq:anisotropic birefringence spectrum}
    \langle \beta^*_{\ell_1 m_1}\beta_{\ell_2 m_2}\rangle&=\langle \beta^{*\text{rec}}_{\ell_1 m_1}\beta^{\text{rec}}_{\ell_2 m_2}\rangle + \langle \beta^{*\text{rei}}_{\ell_1 m_1}\beta^{\text{rei}}_{\ell_2 m_2}\rangle + \langle \beta^{*\text{rec}}_{\ell_1 m_1}\beta^{\text{rei}}_{\ell_2 m_2}\rangle +\langle \beta^{*\text{rei}}_{\ell_1 m_1}\beta^{\text{rec}}_{\ell_2 m_2}\rangle\\
&=\delta_{\ell_1\ell_2}\delta_{m_1 m_2}\left[C^{\beta\beta}_{\ell_1} (\eta_\text{rec}, \eta_\text{rec})+C^{\beta\beta}_{\ell_1} (\eta_\text{rei}, \eta_\text{rei})+2C^{\beta\beta}_{\ell_1} (\eta_\text{rec}, \eta_\text{rei})\right].
\end{align}
Since the cross-term 
\begin{equation}\label{eq:theta PS}
C^{\beta\beta}_\ell (\eta_\text{rec}, \eta_\text{rei})= 4\pi\left(\frac{c_\gamma\alpha_\text{em}}{4\pi }\right)^2 \int\frac{{\rm d}k}{2\pi^2} k^2j_{\ell}(k\Delta\eta_\text{rec})j_{\ell}(k\Delta\eta_\text{rei})P^\theta_{k}(\eta_\text{rec},\eta_\text{rei}),
\end{equation}
is negligible because the two Bessel functions, that peak at $\ell=k \eta$, have little common support because $\eta_\text{rei}\simeq 15 \eta_\text{rec}$, the total power spectrum can be treated as the sum of the contribution from recombination and reionization. We can use this fact, together with the information from the isotropic birefringence, to test different evolutions of the network as we discuss in Sec. \ref{sec:Tomography}.

\begin{figure}
    \centering
    \begin{minipage}[b]{0.5\linewidth}
    \includegraphics[width=7cm]{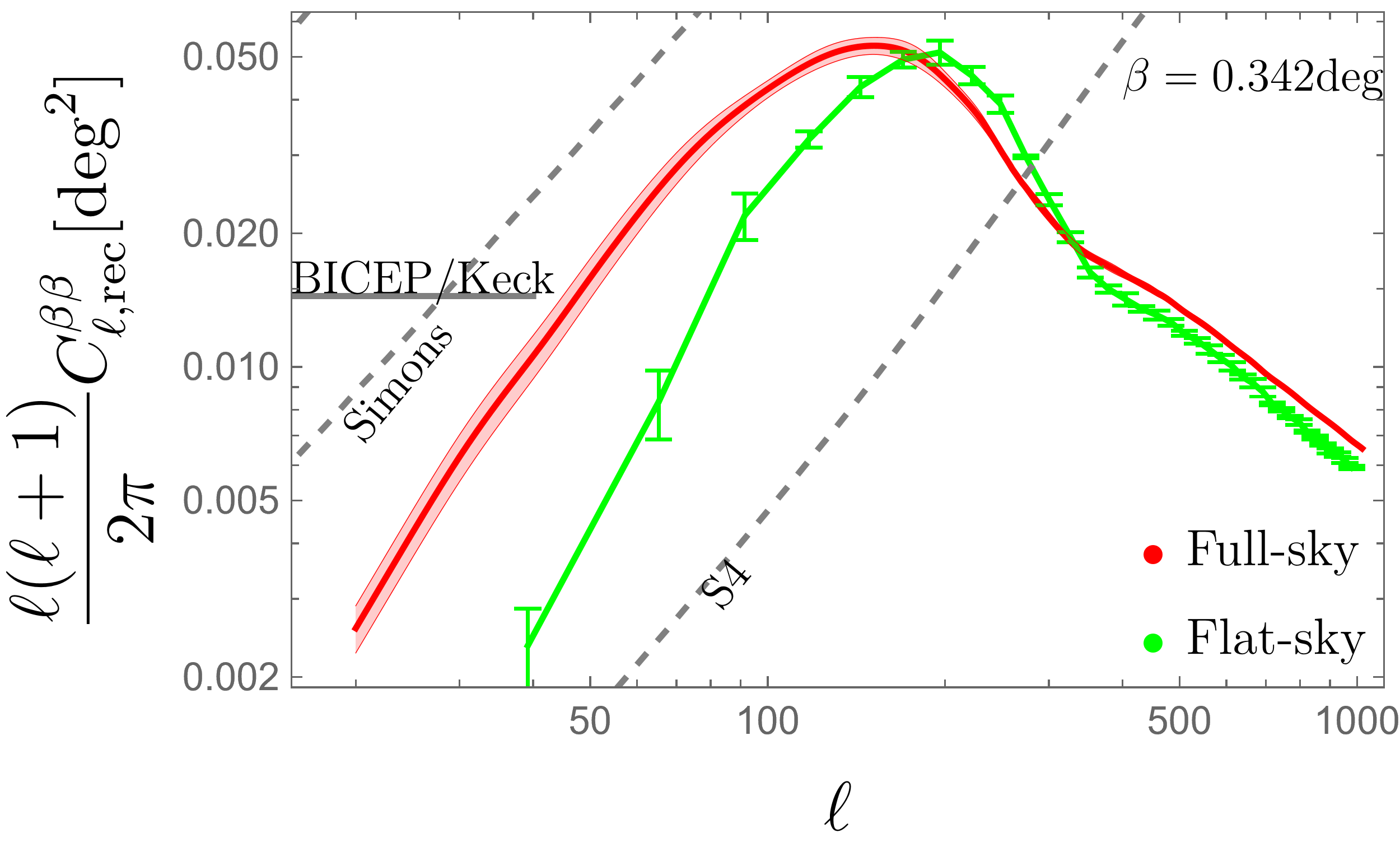}
    \end{minipage}
    \quad
    \begin{minipage}[b]{0.45\linewidth}
    \includegraphics[width=7cm]{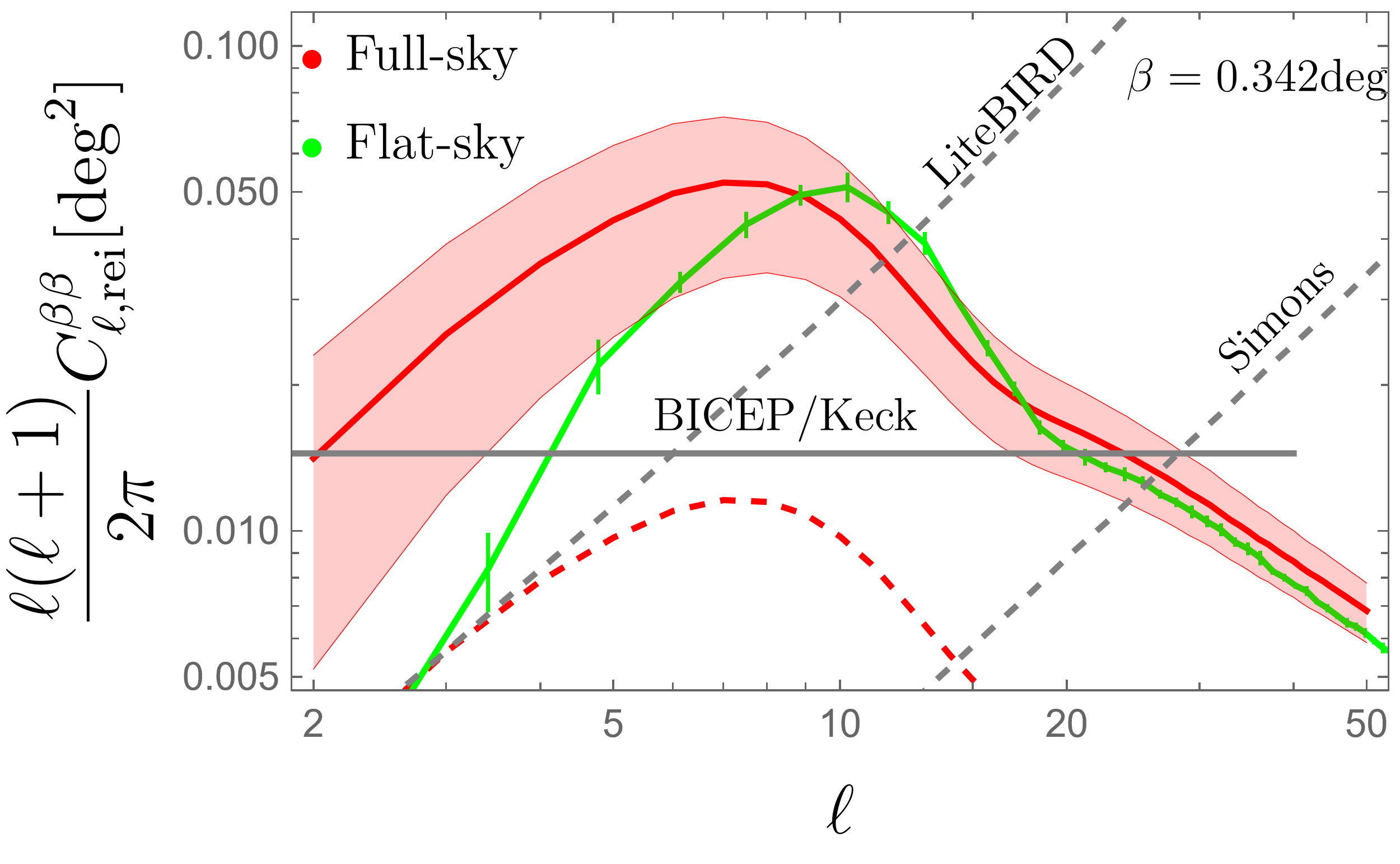}
    \end{minipage}
    \caption{Anisotropic birefringence power spectrum obtained from 3D numerical simulation of DWs computed for recombination (left) and reionization (right). The green line shows the flat-sky result given in eq. \eqref{eq:Flatskyres}, the error bars come directly from those in the power spectrum of Figure \ref{fig:powerspec}. The red lines show the full-sky result computed from eq. \eqref{eq:Cbeta} from the values of the averaged power spectrum shown in Figure \ref{fig:powerspec} with white noise initial conditions. The red error bar on the left panel comes from computing the full-sky result from the one sigma scalar power spectrum of Figure \ref{fig:powerspec} and we checked that the cosmic variance contribution to the error is comparable to this. On the right plot, the error bar is instead mainly given by cosmic variance. The gray lines represent the experimental constraint from BICEP/KECK (solid) \cite{BICEPKeck:2022kci} and the future forecasts with LiteBIRD, Simons Observatory and CMB-S4 (dashed) \cite{Pogosian:2019jbt}.   The BICEP bound looks in tension with the reionization signal for $\beta=0.342$ deg, but both can be reconciled within the two sigma of the isotropic birefringence measurement, i.e. $\beta\simeq 0.16$ deg, as shown by the red dashed line.} 
    \label{fig:Anisotropic birefringence}
\end{figure}

In Figure \ref{fig:Anisotropic birefringence} we show the anisotropic birefringence of the DW network at recombination and reionization computed from eq. \eqref{eq:Cbeta} and compare it with the flat-sky approximation that we derived in Appendix \ref{app:angular PS}. As expected, the anisotropic birefringence peaks at the horizon scale of recombination and reionization, respectively. In the flat-sky approximation, our result closely matches previous results in literature \cite{Takahashi:2020tqv,Kitajima:2022jzz}, thus confirming their robustness. However, we note that on large scales there are sizeable differences between the flat-sky and full-sky results even for the recombination part where we expected those differences to be 
 less pronounced. The full-sky result is numerically more demanding because of the highly oscillating integrand on eq. \eqref{eq:theta PS} so in Appendix \ref{app:FittingCl} we provide some numerical fit for both the recombination and reionization spectra to facilitate future searches.  

\begin{figure}
    \centering
    \begin{minipage}[b]{0.5\linewidth}
    \includegraphics[width=7cm]{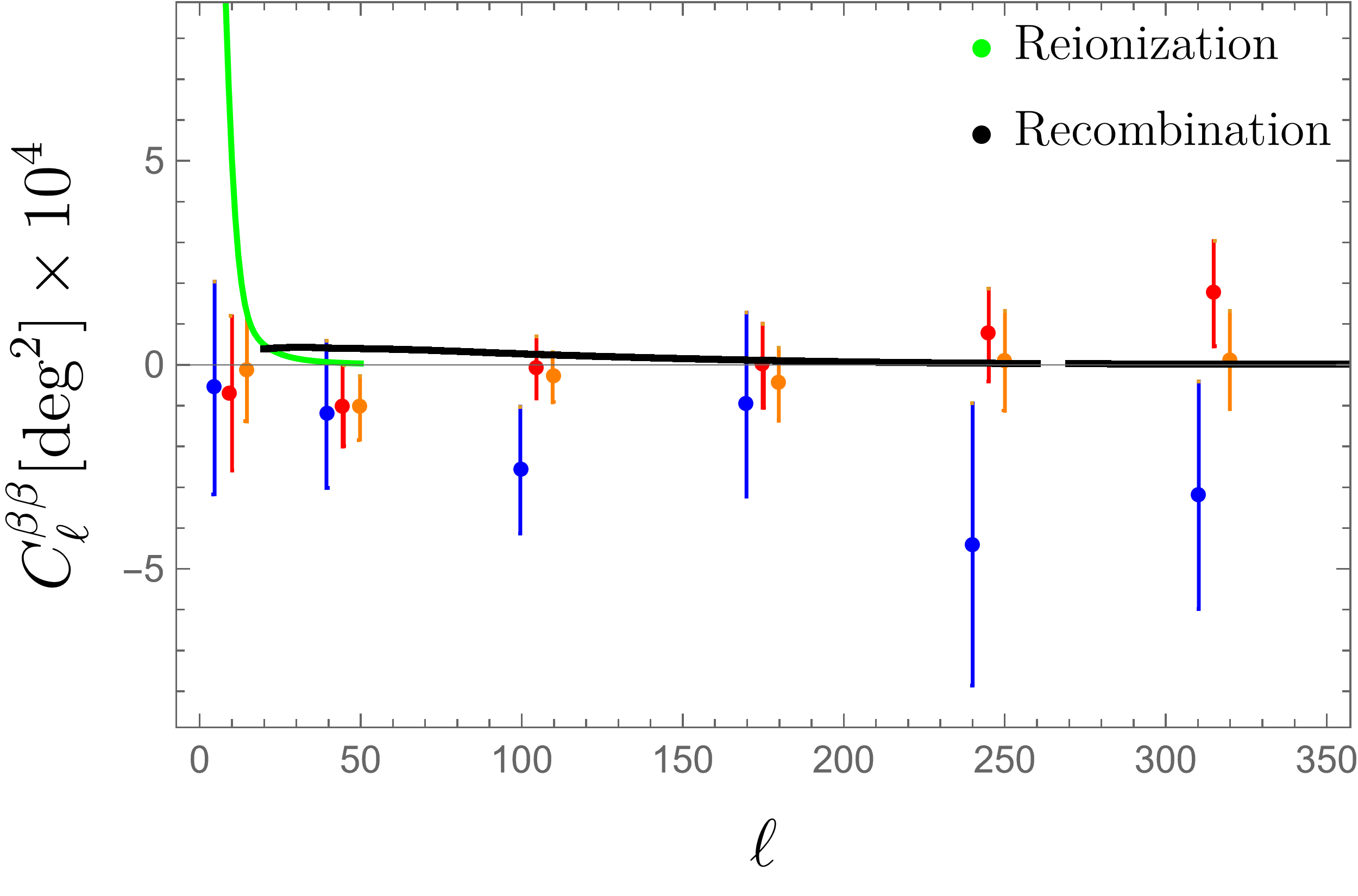}
    \end{minipage}
    \quad
    \begin{minipage}[b]{0.45\linewidth}
    \includegraphics[width=7cm]{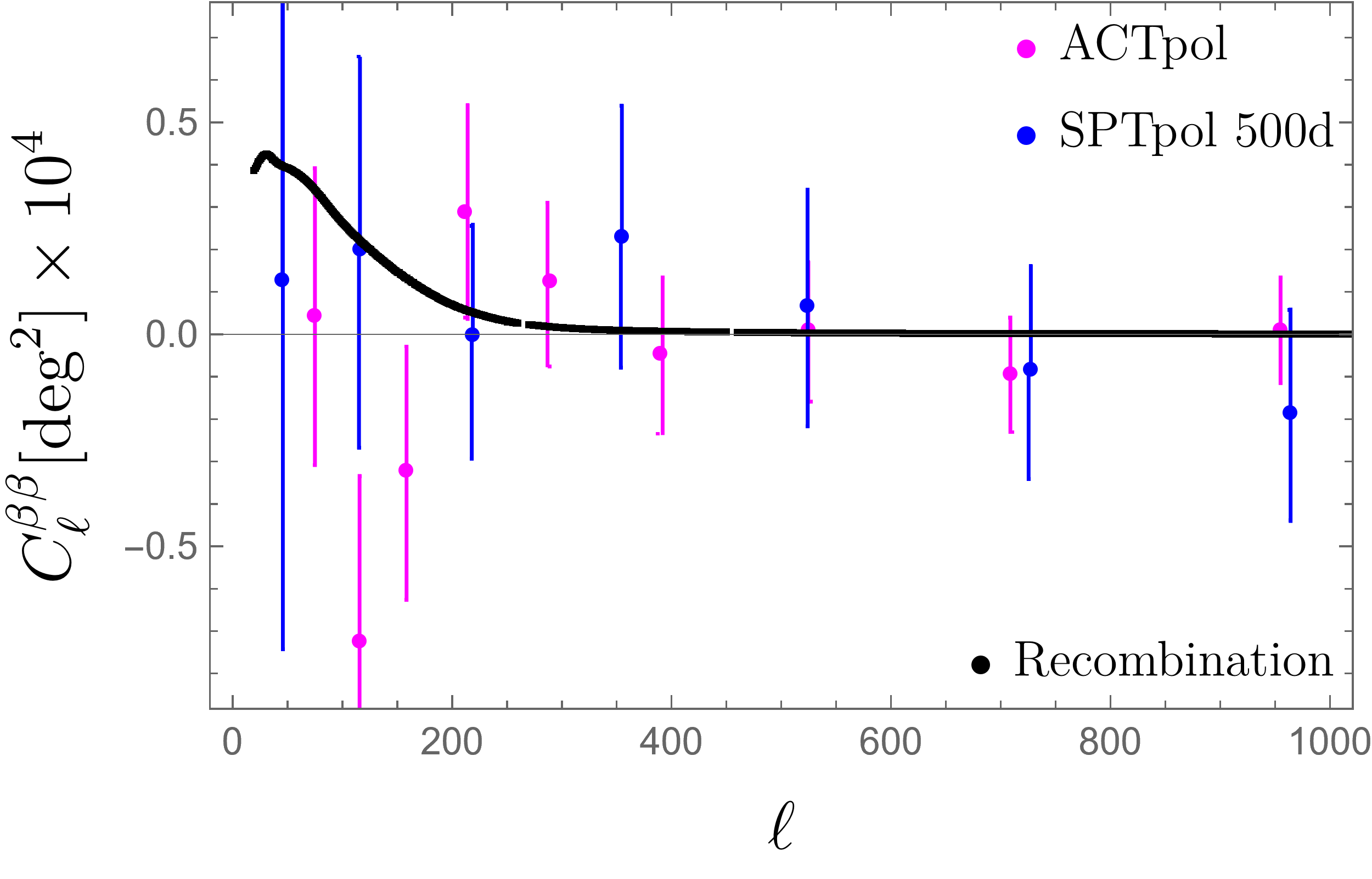}
    \end{minipage}
    \caption{Comparison of $C_\ell^{\beta\beta}$ to observed data, from BICEP/Keck \cite{BICEPKeck:2022kci} on the left and ACTpol \cite{Namikawa:2020ffr} and SPTpol \cite{SPT:2020cxx} on the right. On the left panel, we show both the contribution from recombination and reionization for $\beta\simeq 0.16$ deg, it's clear that data at low multiples already put stringent constraints on the reionization spectra because of its larger amplitude on large scales. As in \cite{BICEPKeck:2022kci}, in the left panel the blue data points correspond to spectra at 150 GHz, red data at 95 GHz and orange data come from the combined spectra. } 
    \label{fig:compBICEP}
\end{figure}

These spectral amplitudes are potentially testable with future CMB measurements, such as Simons Observatory, LiteBIRD and CMB-S4 \cite{Pogosian:2019jbt}.
These sensitivity curves are determined by the statistical uncertainty from beam systematics \cite{Shimon:2007au} whose effect becomes dominant for smaller angular scales. Namely, the anisotropic birefringence power spectrum for the reionization epoch, which has a peak at large angular scales in comparison with that for the recombination epoch, is advantageous for CMB measurements. Indeed, by comparing our prediction to current constraints on the scale-invariant power spectrum from BICEP/Keck \cite{BICEPKeck:2022kci}, it looks like there is already a tension between the two. This can however be relaxed by lowering the isotropic signal at reionization within the two-sigma range $\beta \rightarrow 0.342-2\sigma\simeq 0.16$ deg as shown with the red dashed line on the right panel of Figure \ref{fig:Anisotropic birefringence}.
To have a more direct comparison with the observed data and the sensitivity over different multipoles in Figure \ref{fig:compBICEP} we compare our spectra with the data of BICEP/Keck \cite{BICEPKeck:2022kci} on the left and of  ACTpol \cite{Namikawa:2020ffr} and SPTpol \cite{SPT:2020cxx} on the right.  

\subsection{Effects of the finite thickness of the last scattering surface and axion mass}

Some comments are in order: first, neither recombination nor reionization are instantaneous during the evolution of the Universe thus their thickness could leave some effect on the observed anisotropies; second, the walls move in random directions at relativistic speeds which can also affect the observed spectrum. 
Regarding the finite width of the last scattering surface, which we denote as $\Delta \eta_{\rm rec}$,
one would need to convolute the scalar power spectrum with the visibility function, but we believe this to be a small correction as we now explain. The scalar power spectrum always  peaks at the same comoving scale $k\sim 1/\eta$, thus, we expect the finite width to smear the peak of the signal over the scales
\[\ell_{\rm peak}\simeq\frac{\eta_0}{\eta_{\rm rec}\pm \Delta \eta_{\rm rec}}\simeq\frac{\eta_0}{\eta_{\rm rec}} \left(1\mp \frac{ \Delta \eta_{\rm rec}}{ \eta_{\rm rec}}\right).\] Given that $  \Delta \eta_{\rm rec} /\eta_{\rm rec}=0.04$, this is a small effect at recombination and it changes the value of the peak by $\Delta\ell_{\rm peak}\sim \mathcal{O}(1)$ at $\ell_{\rm peak}\sim100$. The effect is however more important at reionization since the relative thickness of this period is bigger.  Cosmic variance, estimated by $\Delta C_\ell^{\beta\beta}\simeq \, \sqrt{2/(2\ell+1)} C_\ell^{\beta\beta}$ \cite{Takahashi:2020tqv}, is another source of uncertainty at these multipoles which impacts the reionization spectrum as shown in the error bars of Figure \ref{fig:Anisotropic birefringence}.

Regarding the relativistic velocities of the DWs, we expect their relative motion to lead to some fuzziness in the anisotropies at scales smaller than the typical distance\footnote{The typical velocity of the DWs is close to the speed of light. To be conservative, we take $ \sqrt{\langle v^2_{\rm DW}\rangle}\sim c$ which gets reduced by a factor of two if one assumes that the orientation of the DW with respect to the line of sight, denoted by $\Theta$, is random and isotropic $\langle \sin(\Theta)\rangle=\int_0^{\pi/2}\cos(\Theta)\sin(\Theta){\rm d}\Theta=0.5$.} travelled by the walls during recombination 
 $ \Delta \eta_{\rm rec}$. Taking as typical DW velocity projected on the sky $0.5 \sqrt{\langle v^2_{\rm DW}\rangle}$, the corresponding angular scale is
\begin{equation}
 \Delta L_{\rm DW} \sim 0.5 c{ \Delta \eta_{\rm rec}}\sim 0.02 \eta_{\rm rec} \qquad \rightarrow \qquad \theta\simeq \frac{\Delta L_{\rm DW} }{\eta_0}\simeq 0.02\frac{ \eta_{\rm rec} }{\eta_0}\simeq 0.02 \, {\rm deg},
 \end{equation}
which corresponds to large multipoles
 $\ell_{\rm fuzz} \sim \frac{\eta_0}{\Delta L_{\rm DW}}=  50\frac{\eta_0}{\eta_{\rm rec}}\simeq 2500$. Thus, we do not expect these effects to significantly modify the birefringence spectrum shown in Figure \ref{fig:Anisotropic birefringence}, nor the expectation for the isotropic birefringence which is an averaged quantity over the sky thus small-scale variations are not important.
 
 Lastly, we comment on the effect of the mass on the spectrum. Because the scalar power spectrum decays exponentially on scales $k\gtrsim k_m=1/m_\phi$, the spectrum shown in Figure \ref{fig:Anisotropic birefringence} will be suppressed at multipoles larger than 
\begin{equation}
    \ell_m\sim \frac{k_m}{10^{-4}}\frac{\rm Mpc}{h_0}\sim \frac{100}{h_0} \left( \frac{10^{-27} {\rm eV}}{m_\phi}\right) \, .
\end{equation}
This suppression of anisotropies will be pronounced if the axion mass is smaller or comparable to the Hubble rate at recombination.

\section{Tomographic probes of annihilating and late-forming  networks}
\label{sec:Tomography}

Recent analyses of the CMB data have found evidence for a non-vanishing signal of $\beta(z_\text{rec})=0.342^{+0.094}_{-0.091}$ deg at 68\% CL \cite{Eskilt:2022cff}.
However, the amount of rotation can in principle vary in time and recent works proposed a way to probe the time dependence through a tomographic analysis. 
The polarization of the CMB photons is mostly generated at two characteristic times:  recombination ($z_\text{rec}\sim 1100$, from Thomson scattering and quadrupole anisotropies), and reionization ($z_{\rm rei}\sim 8$, when the first stars reionize the intergalactic medium) \cite{Planck:2018vyg}.
Therefore, we expect two peaks in the $C_l^{EB}$ spectrum at the two corresponding angular scales. A possible detection of such peaks with future data would then allow to distinguish the amount of birefringence generated from recombination from that generated from reionization \cite{Sherwin:2021vgb,Nakatsuka:2022epj, Galaverni:2023zhv}\footnote{Polarized Sunyaev-Zel'dovich effects by the scattering of CMB photons by electrons in galaxy clusters are potentially another tomographic observable that could allow to test the cosmic birefringence at redshifts lower than the cosmic reionization epoch \cite{Lee:2022udm,Namikawa:2023zux}.}.
To make contact with the toy model of ref. \cite{Nakatsuka:2022epj}, where it is studied the impact of a redshift dependent isotropic signal on $C_\ell^{EB}$, it is convenient to simplify the redshift dependence of $\beta$ as \cite{Nakatsuka:2022epj}:
\begin{equation} \label{eq: redshift dependent beta}
	\beta(z)=\begin{cases}
		0 & , \, z=0  \\
		\beta_{\text{rei}} & , \, 0<z\leq z_{\rm rei}\\
		\beta_{\text{rec}} & ,\, z> z_{\rm rei} \, .
	\end{cases}
\end{equation}

\begin{figure}[t]
    \centering
\includegraphics[width=0.47\linewidth]{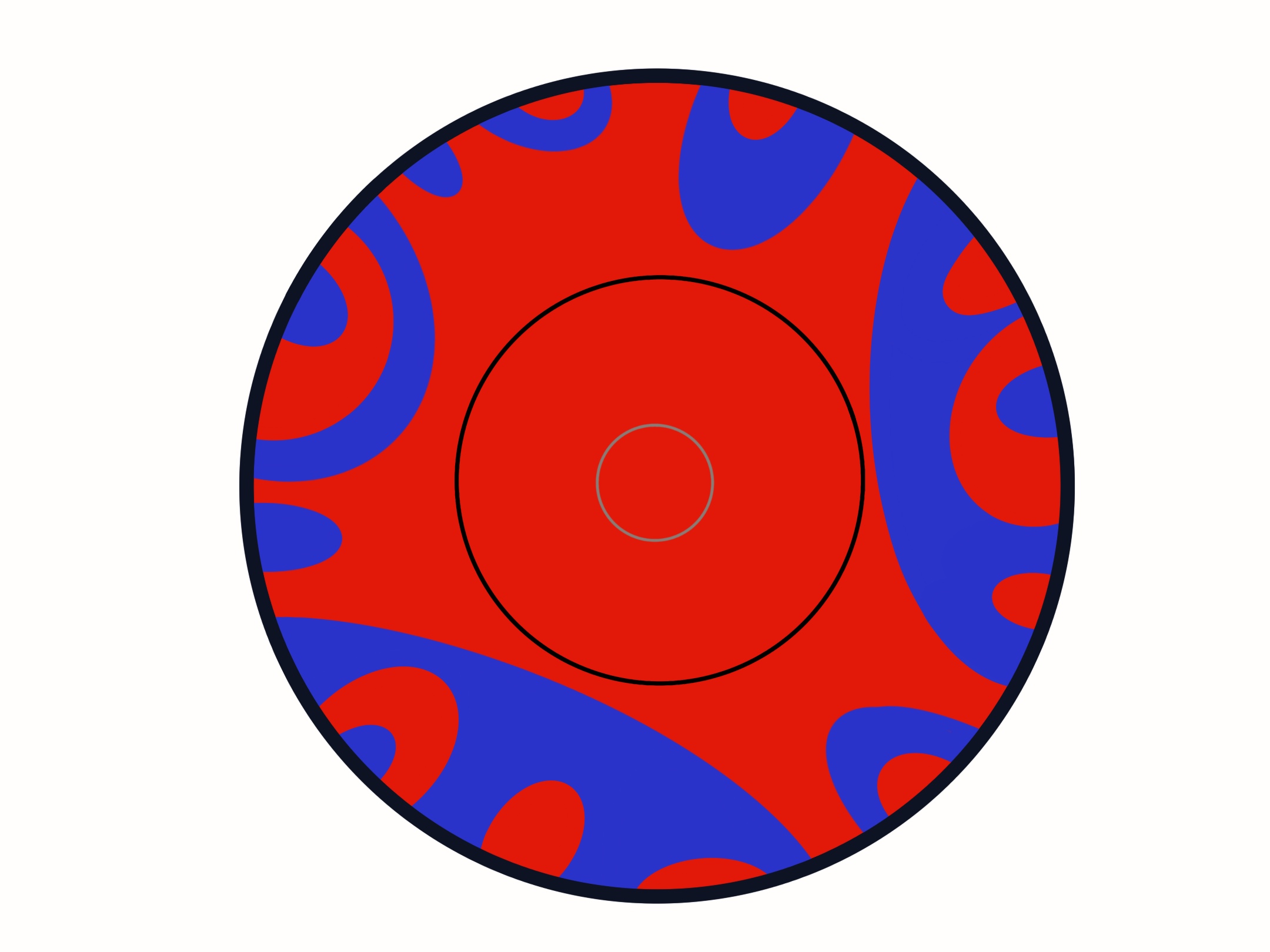}
    \caption{ 
    ``Celestial plot'' representation of an annihilating DW network in a $Z_2$ model (only 2 vacua). The  annihilation redshift corresponds to the solid black circle. From then on, all patches are in one and the same vacuum.
    Note that the contribution of the isotropic birefringent angle from recombination would coincide with the one from a network in scaling (depicted in Fig.~\ref{fig:sketchZ2}), in magnitude. In the annihilating network case, the {\em sign} of the signal can be predicted.   
    }
    \label{fig:sketchAnn}
\end{figure}

There are several distinct cases, that we summarize in Table \ref{table II}.
When the network forms before recombination and survives until today one expects the average value of $\theta$ at recombination and reionization to be similar $\left<\theta_\text{rei}\right>\simeq \left<\theta_\text{rec} \right>$ \cite{Takahashi:2020tqv,Kitajima:2022jzz} \footnote{The average value of $\langle \theta (z) \rangle$ is computed by summing over the number of Hubble patches $N_{\rm patches}$  at redshift $z$ and so has fluctuations that decrease with $1/\sqrt{N_{\rm patches}}$. At recombination there are $N_{\rm patches}\simeq 7\times 10^3$ then the relative uncertainty on the isotropic birefringence is $\sigma_\beta/\langle\beta\rangle\simeq 0.01$; at reionization, the number of patches is much smaller $N_{\rm patches}\simeq 30$ which translates into a relative uncertainty of  $\sigma_\beta/\langle\beta\rangle\simeq 0.18$. Overall this is a small error and we will neglect it in the rest of the paper.}.
However, that might no longer be the case when the network forms or annihilates between recombination and the present time as depicted in Figure \ref{fig:sketchAnn}. 
If the network forms after recombination the value of  $\left<\theta_\text{rec}\right>$ can change depending on the details of the scalar field potential at that time - before the defects are formed.

If, instead, the network annihilates after recombination the prediction for $\beta_\text{rec}$ remains essentially the same in the case of the $Z_2$ DW network, with the only difference being that the local value $\theta_\text{loc}$ would then correspond to one single global minimum and the final result becomes deterministic. In the case of the string-wall network, the environmental effect from local structures is suppressed for large $z_\text{ann}$. However, there can still be a leftover signal for example from the fact that the misaligned contribution to the axion potential the one that can cause the annihilation of the network, is in general not aligned with the potential $V$ in eq. \ref{pot} \footnote{For example, if the misaligned potential $V_\text{mis}$,  moves the global minima to a field value that is asymmetric with respect to the minima of $V$ then photons that come from the different minima of $V$, while the network was in scaling, travel through different field excursions thus leaving again a signal on the ballpark of $\beta_\text{rec}\sim 0.21 c_\gamma/N$ deg.}.
In any case, if the annihilation happens before reionization, there will be no further rotation of the photon polarization axis after that point implying that $\beta_\text{rec}\neq \beta_\text{rei}=0$. 
The tomographic measurement of the isotropic birefringence as proposed in \cite{Sherwin:2021vgb,Nakatsuka:2022epj} can then allow to measure $\beta_\text{rei}$ and $\beta_\text{rec}$ and so distinguish between scenarios with different annihilation times, although not between scenarios with different formation times. For the latter, the information stored in the anisotropic signal, described in the next paragraphs, can provide further input as summarized in Table \ref{table II}.

\setlength{\arrayrulewidth}{0.2mm}
\renewcommand{\arraystretch}{2}
\setlength{\tabcolsep}{1.2pt}
  \setlength\extrarowheight{-3pt}
\begin{table}[]
\centering
\vspace{0.2cm}
\begin{tabular}{|l||*{2}{c|}}\hline
\backslashbox{\text{\textbf{Formation}}}{\text{\textbf{Annihilation}}}
&\makebox[8em]{$z_{\text{rec}}>z_{\text{ann}}>z_{\text{rei}}$} &\makebox[8em]{ $z_{\text{rei}}>z_{\text{ann}}\geq 0$ }\\\hline 
\hline
\textit{Isotropic Birefringence:} & & 
\\
\hline
\, \,
& {$\beta_\text{rec}\neq \beta_{\rm rei}=0$}&{$\beta_\text{rec}\simeq\beta_{\rm rei}\neq0$} \\\hline \hline
\textit{Anisotropic Birefringence:} & & 
\\
\hline
\, \, $z_{\text{f}}>z_{\text{rec}}$  &$C_{\rm DW}^{\beta\beta}|_{\text{rec}},\quad C_{\rm DW}^{\beta\beta}|_{\text{rei}}\sim0$   & $C_{\rm DW}^{\beta\beta}|_{\text{rec}},\quad C_{\rm DW}^{\beta\beta}|_{\text{rei}}$\\\hline
\, \, $z_{\text{rec}}>z_{\text{f}}>z_{\text{rei}}$ &$C_{\text{IC}}^{\beta\beta}|_\text{rec},\quad C_{\rm DW}^{\beta\beta}|_{\text{rei}}\sim0$      &    $C_{\text{IC}}^{\beta\beta}|_\text{rec},\quad C_{\rm DW}^{\beta\beta}|_\text{rei}$ \\\hline
\, \, $z_{\text{rei}}>z_{\text{f}}>0$    &              //                            &   $C_{\text{IC}}^{\beta\beta}|_\text{rec},\quad C_{\text{IC}}^{\beta\beta}|_\text{rei}$ \\\hline
\end{tabular}
\caption{Properties of the cosmic isotropic and anisotropic birefringence caused by DW networks with different times of formation and annihilation. The isotropic value depends only on the annihilation time whereas the anisotropic component depends on both. We distinguish when the anisotropic spectrum is determined by the DW network $C^{\beta\beta}_{\rm DW}$ from when it is sensitive to the initial condition of the scalar field $C^{\beta\beta}_{\rm IC}$. We expect that most of these features can be extrapolated to the case of string and string-wall networks.}
\label{table II}
\end{table}

Let's first consider the case where the network does not annihilate or annihilates after reionization  (second column of Table \ref{table II}). 
If the network has formed before recombination the anisotropic signal would show a double peak spectrum, corresponding to the sum of the spectra of recombination and reionization as shown in Figure \ref{fig:Anisotropic birefringence}. In case the network is not yet formed at the time of recombination, the anisotropies at those scales will likely be much smaller and sensitive to the initial conditions of the scalar field before the walls are formed $C^{\beta\beta}_{\rm IC}$. In the remaining case, when the network forms after reionization, the phenomenology will be hard to distinguish from the axion-like case with masses $m_\phi\lesssim 10^{-32}$ eV because the anisotropies at recombination and reionization are both small. 

If instead, the network annihilates between recombination and reionization (first column of Table \ref{table II}).
Then, both the isotropic and anisotropic reionization components, $\beta_{\rm rei}$ and $C^{\beta\beta}|_{\rm rei}$, vanish. If, moreover, the network is not yet formed at the time of recombination, the anisotropies at those scales will again also likely be small and initial condition dependent. 
We expect that most of these features can be extrapolated to the string and string-wall networks with the caution of adapting $C^{\beta\beta}|_{\rm rec,rei}$ to the presence of the string loop effects that can affect the overall spectrum.

So even if future analyses do not detect anisotropies on the birefringence angle a late-formed domain wall network is not ruled out but it could still be probed by other observational signatures, such as the effect on polarized sources at low redshifts like Quasars \cite{Sluse:2005bg} or Pulsars \cite{Liu:2021zlt,Castillo:2022zfl}. Interestingly, all the previously discussed scenarios can leave an imprint on the B-modes of the CMB as we discuss in the next section.

\section{Gravitational waves from networks with DWs}
\label{sec: GWs}

The axionic networks are very inhomogeneous on scales around the horizon, i.e. $\delta \rho(k_H, \eta) \sim \rho(\eta)$, where $\rho$ is the network's energy density and $\delta \rho (k_H, \eta)$ its perturbations from the average value at the horizon scale $k_H$ at the time $\eta$. Therefore, if present at CMB times the network can leave sizeable effects on the CMB perturbations even when not coupled to photons.

The gravitational effect of strings in the CMB is typically small unless the string tension is considerably large (see e.g. \cite{Rybak:2021scp} for a recent analysis). However, the situation changes dramatically when DWs are present (with or without strings) due to their growing abundance, as discussed in \ref{sec: DW networks}. In this latter case, previous works have focused on the CMB effects of networks that: i) remain stable until today, $\sigma^{1/3} \lesssim$ MeV \cite{Zeldovich:1974uw,Lazanu:2015fua,Sousa:2015cqa}, ii) annihilate before recombination \cite{Ramberg:2022irf}. Here, we extend these works by studying the gravitational waves (GW) generated by the network while in scaling and their impact on the CMB B modes, and by considering networks that annihilate between recombination and today.  To this end, we have used the dedicated numerical simulations, described in appendix \ref{app: Numerical Simulation}, to derive the GW spectrum of the DW network which we will then use to derive bounds on the wall tension $\sigma$ (or equivalently $\Omega_\text{ann}$) using existing constraints on stochastic GW backgrounds at CMB scales \cite{Namikawa:2019tax}.
\begin{figure}
	\centering
	\includegraphics[width=0.7\linewidth]{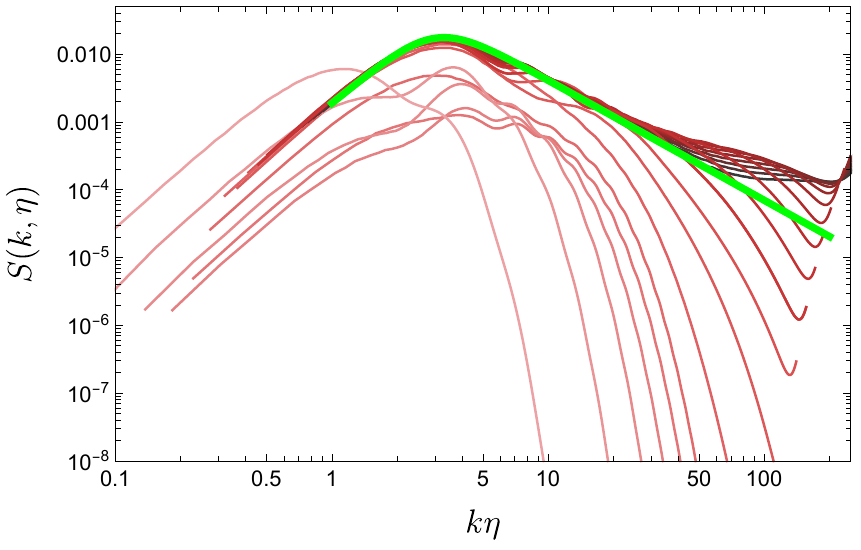}
	\caption{Time evolution of the spectral function $S(k,\eta)$ (from light red to dark red) as a function of  $k \eta$ and for thermal initial conditions. The green curve is the fit to the latest time of the simulation. The data points and their errors are not shown in this figure to avoid cluttering. See Appendix \ref{app: Fitting the GW spectrum} for more details.}
\label{fig:plotm41agwrescaled}
\end{figure}
While in the scaling regime, the network abundance  $\Omega_\text{dw}(\eta)$ in eq. \ref{eq:DW abundance}  grows in time and so does its GW spectrum that is proportional to $\Omega_\text{dw}^2$ \cite{Hiramatsu:2013qaa}. Therefore, the GW signal is expected to peak around the time the network annihilates.~\footnote{Note that after annihilation the network will decay into its massive scalar field components. Therefore, the scalar field abundance will remain approximately constant after annihilation (unless it decays into radiation in the meantime). However,  in that regime,  the GW spectrum moves towards smaller, non-CMB, scales as there are no longer coherent structures of cosmological size.} We follow the approach used in previous works and estimate the GW signal from the network to be that of the scaling regime evaluated at the time of annihilation.

In terms of the microphysical parameters of the DW, the gravitational wave spectrum $\Omega_\text{GW}(\eta,k)$ is characterized by \cite{Hiramatsu:2013qaa}
\begin{eqnarray} \label{eq: GW spectrum}
	\Omega_\text{GW}(\eta,k) = \frac{1}{\rho_\text{tot}(\eta)} \frac{d \rho_\text{GW}(\eta,k)}{d\log k}\,  =    \parfrac{\rho_\text{dw}}{\rho_\text{tot}}^2    S(k, \eta) \,.
\end{eqnarray}
where  $S(k,\eta)$ is a spectral function that in the scaling regime reaches an attractor shape that only depends on the product $k \eta$, 
\begin{eqnarray}
	S(k,\eta)\to S_{sc}(k \eta) \, ,
\end{eqnarray}
and peaks at the horizon scale. We show in Fig. \ref{fig:plotm41agwrescaled} the time-evolution of $S(k,\eta)$ throughout the simulation (where the comoving momenta $k$ are kept fixed). The parameters describing the simulations are described in Appendix \ref{app: Numerical Simulation}. The spectrum evolves from early times (light red), where we have chosen thermal initial conditions for the scalar field, to the scaling regime at late times (dark red) where the spectrum reaches the attractor shape  $S_{sc}(k \eta)$. The case with white noise initial conditions is shown in Fig. \ref{fig:scaledgwspectrum-error-thermal-l80}.

To reconstruct the shape function we have taken the scalar field data at the latest time of the simulation \footnote{To find the fitting function we discarded the data at the smallest scales, as their evolution is influenced by the finite width of the wall.} in Fig. \ref{fig:plotm41agwrescaled}  (taking into account the 1$\sigma$ errors obtained after averaging over 5 realizations) and fitted the spectral shape $S_{sc}(k \eta)$ to the double power-law function
\begin{eqnarray} \label{eq: Fitting formula}
	S_\text{sc}^\text{fit} (x)= A \frac{\theta+\beta}{\theta \parfrac{x}{x_{peak}}^{-\beta}+\beta \parfrac{x}{x_{peak}}^\theta} \, .
\end{eqnarray}
The best fit is shown in green in Fig. \ref{fig:plotm41agwrescaled} and corresponds to the best-fit parameters: $A \simeq0.02,\beta \simeq 2.9,\theta\simeq 1.7$ and $x_{peak}\simeq 3$ (see Appendix \ref{app: Fitting the GW spectrum} for more details about the fit and the data selected). We find this functional form of the fit to be a good description of the data for $k\eta \lesssim 50$. For even smaller scales we see the appearance of a flattening region, in the case of thermal initial conditions, or a bump in the case of white noise initial conditions. It is unclear at this point if this behavior is caused by the finite width of the wall, by some remnant of the initial conditions that have not yet diluted, or if it is a physical feature. Therefore, we discard the data points with $k \eta >50$ when performing the fit. A dedicated study with a larger numerical simulation should clarify the origin of the change of slope at those scales. 

Let us now discuss the best-fit parameters obtained.
The position of the peak at 
$k\eta \simeq 3$ is compatible with the expectation of a peak at the scale of the horizon $k_\text{peak} \sim a H \sim 2/ \eta$ where the factor of 2 is present for a matter-dominated background evolution. We find  $\beta=4.3$ for the slope of the spectrum at low momenta, stronger than that dictated by the causality argument ($\propto k$ in a matter-dominated regime). Note, however, that the causality slope is only expected once scales that were superhorizon at the time the signal was generated reenter the horizon, and not for these intermediate horizon-size lengths. On the other hand, for high momenta, at the right of the peak, we find a slope of $\theta=1.7$. \footnote{Including more UV points in the fit moves the best fit to values closer to one, see discussion in App. \ref{app: Fitting the GW spectrum}.} This value is larger than the value obtained in previous studies in radiation domination backgrounds where $\theta$ was found to be around one \cite{Hiramatsu:2013qaa}. The best-fit values for the case with white noise initial conditions are shown in Table \ref{table: best fit values}.

The attractor nature of $S_{sc}(k \eta)$, allows to extrapolate the spectrum of GWs obtained from our numerical simulation to other networks with different DW tensions and annihilation times, by means of eq. \ref{eq: GW spectrum} and the appropriate rescaling of $\rho_\text{dw}$ and $\rho_\text{tot}$.~\footnote{The annihilation of the network is a violent process that includes the collapse of closed domains and thus can further source gravitational waves (see e.g. \cite{Kitajima:2023cek,Chang:2023rll,Li:2023gil}). Here we neglect these potential additional sources of GWs. Our approach is, in this sense, conservative.} 
The next step is to evolve the spectrum from annihilation to the present time. 
The wave numbers that are subhorizon at the time of annihilation redshift as radiation, and so faster than the matter-dominated background by a factor of $a$, the scale factor. For these modes, the GW spectrum today is given by
\begin{figure}[t]
	\centering    
    \includegraphics[width=0.7\textwidth]{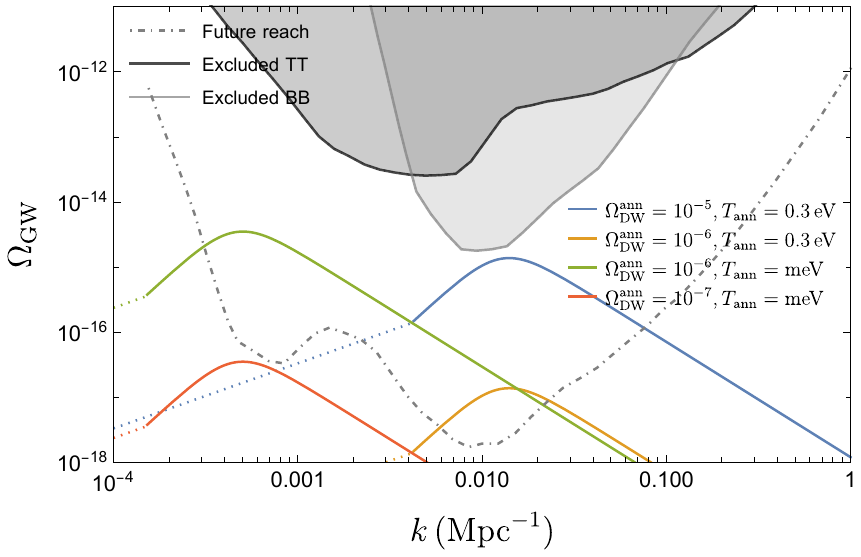}
	\caption{Fitted GW spectrum (thermal initial conditions) from DW networks with four different sets of parameters (see main text for more details).  Black and gray regions correspond to current CMB constraints on a stochastic spectrum of GWs, and the region above the dot-dashed line are the prospects with future CMB experiments \cite{Namikawa:2019tax}. 
	}
	\label{fig:OmegaGWplot}
\end{figure}
\begin{eqnarray} 
	\label{eq:GW spectrum today}	
	\Omega_\text{GW}(\eta_0,k) &=&  \parfrac{a_\text{ann}}{a_0}\, \left(\Omega_\text{dw}^\text{ann}\right)^2 	 S_{sc}\left(k \eta_0\right)
\end{eqnarray}
where $\Omega_\text{dw}^\text{ann}\equiv \Omega_\text{dw}(\eta_\text{ann})$ and $a_\text{ann}/a_0 \simeq T_{0}/T_\text{ann}$ with $T_0$ ($T_\text{ann}$) the CMB temperature at the present  (annihilation) time.  Regarding the modes that are superhorizon at $\eta_\text{ann}$, we extrapolate their GW spectrum using the causality argument that sets the slope of the spectrum to be $k$ once they re-enter the horizon (rather than the $k^3$ for radiation domination) \cite{Cai:2019cdl}. 
Finally, 
we use the approximate expression $\eta \simeq 2/(\sqrt{\Omega_m} H_0) \left(\sqrt{a+a_\text{eq}} -\sqrt{a_\text{eq}}\right)$, valid for a universe filled with matter and radiation~\footnote{The correction from dark energy is small, of approximately 20\%.}, to relate the annihilation time $\eta_\text{ann}$ with the present conformal time $\eta_0$, where $\Omega_m,H_0$ are respectively the matter abundance and the Hubble rate today that we take from the latest Planck analysis \cite{Planck:2018vyg}. 

In Fig. \ref{fig:OmegaGWplot} we show the resulting GW spectrum as a function of the physical momenta today for different values of $\Omega_\text{dw}^\text{ann}$ and the annihilation temperature $T_\text{ann}$ using the fitted spectral function discussed above (with thermal initial conditions). The dotted lines correspond to the extrapolation of the spectrum to the scales that were superhorizon at $\eta_\text{ann}$. For fixed $\Omega_\text{dw}^\text{ann}$, networks that annihilate later have a  larger  GW spectrum because the GWs redshift faster than the background for a shorter time. On the other hand, different annihilation temperatures move the peak of the spectrum to different scales. 
In the same plot, we also show the current CMB bounds on a  general stochastic spectrum of GWs, that we will use in the next paragraphs to constraint the parameters of the model, and the prospects for detection with future CMB experiments   \cite{Namikawa:2014lla,Namikawa:2019tax}.

\begin{figure}[t]
	\centering    
\includegraphics[width=0.485\linewidth]{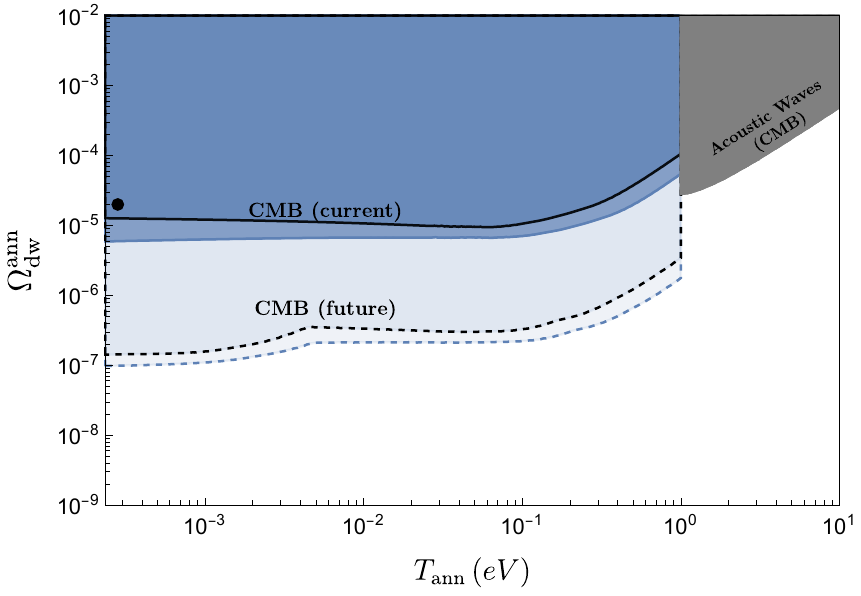}\, \,
	\includegraphics[width=0.485\linewidth]{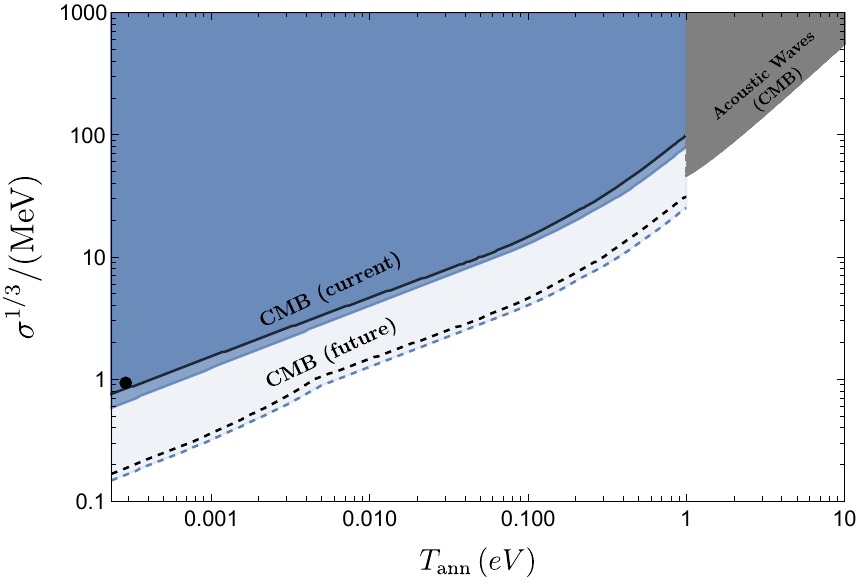}
	\caption{Left: Bounds on the DW abundance at the time of annihilation of the network $\Omega_\text{dw}^\text{ann}\equiv \rho_\text{dw}/\rho_\text{tot}(T_\text{ann})$ as a function of the CMB temperature at the same time (this work). The dark blue region is excluded by current CMB data due to the overproduction of GWs. The light blue region shows the prospects for detection with an idealized future CMB experiment \cite{Namikawa:2019tax} (see main text for more details). Black contours correspond to the same bounds but with white noise (rather than thermal) initial conditions.
 		The gray region is excluded by the CMB analysis of \cite{Ramberg:2022irf} and the black circles denote the bounds on stable DW networks ($T_\text{ann} \lesssim 2.7$K) from \cite{Zeldovich:1974uw,Lazanu:2015fua,Sousa:2015cqa}. Right: Same as the left plot but in terms of the wall tension.}
	\label{fig:Bounds on OmegaGW}
\end{figure}

We proceed with a systematic analysis where we use the GW spectrum in eq. \eqref{eq:GW spectrum today} to search for the region of parameters in the $\Omega_\text{dw}^\text{ann}$--$T_\text{ann}$ plane where the spectrum overlaps with the region excluded by the CMB bounds derived in \cite{Namikawa:2019tax} and shown in Fig. \ref{fig:OmegaGWplot}. We show the resulting constraints in Fig. \ref{fig:Bounds on OmegaGW} in terms of the DW abundance $\Omega_\text{dw}^\text{ann}$ (left) and DW tension $\sigma^{1/3}$ (right).~\footnote{To obtain the bounds for the tension of the wall $\sigma$, right plot of Fig. \ref{fig:Bounds on OmegaGW},  we have used the scaling relation $\rho_\text{dw}=c \sigma H$ with $c=3/2\cdot 0.5$ that we take from the numerical simulations with thermal initial conditions (the case of white noise is only slightly different with $c=3/2\cdot 0.8$). On the other hand,  to obtain the gray region in the left plot, where we show the bounds derived in \cite{Ramberg:2022irf} that apply for DWs that annihilate in a radiation-dominated regime, we have instead used $c=2\cdot 0.8$ \cite{Hiramatsu:2013qaa} to convert the bounds into $\Omega_\text{dw}^\text{ann}$.} The dark blue region is excluded by the current data 
  whereas the region in light blue could be detected by an idealized full sky CMB experiment with $\mu$K-arcmin white noise with an arcmin Gaussian beam \cite{Namikawa:2019tax}.~\footnote{An important observation is that we are considering axion masses $m_\phi\gg H_{\rm rec}$ so that the suppression of the spectrum at scales $k \gg m_\phi^{-1}$ lies outside the scales probed by CMB. The suppression occurs at scales higher than $k\gtrsim m_\phi^{-1}\sim 0.01{\rm{Mpc}} \, (10^{-27}\rm{eV}/ m_\phi)$.} The black (continuous and dashed) contours indicate the same constraints but  using instead the GW spectrum from the simulation with white noise initial conditions (instead of thermal). We note that the constraints are not very sensitive to this dependence.
Finally, in gray we show the CMB constraints on DW networks that annihilate before recombination  \cite{Ramberg:2022irf}. 

Roughly speaking, the current CMB data limits $\Omega_\text{dw}^\text{ann}$ to values smaller than $10^{-4}-6\cdot 10^{-6}$ and DW tensions below $0.5-100$ MeV, depending on the annihilation temperature, and future CMB experiments will in principle be able to probe DW networks with abundances down to $10^{-7}$ and tensions down to $0.1$ MeV. The bound on $\Omega_\text{dw}^\text{ann}$ is in the ballpark of what one would expect by roughly comparing $\Omega_\text{dw}^\text{ann}$ with the amplitude of the CMB anisotropies.  Note also that even if the CMB bounds shown in Fig. \ref{fig:OmegaGWplot} are peaked at $k\sim 0.01 \text{Mpc}^{-1}$, the constraints on $\Omega_\text{dw}$ are rather flat on $T_{\rm{ann}}$.  
This is because of the shallower spectral shape of the GW signal and the fact that the GW peak for networks that annihilate after recombination lies at $k\lesssim 0.01 \text{Mpc}^{-1}$.

Our analysis also extends to networks that survive until today, with the GW spectrum given in that case by eq. \ref{eq:GW spectrum today} with $T_\text{ann} \simeq T_\text{CMB} \simeq 2.7$ K. We find  the resulting bounds to be stronger than in previous analyses  (black circles in Fig. \ref{fig:Bounds on OmegaGW}) \cite{Lazanu:2015fua,Sousa:2015cqa}. We believe that this improvement is justified by the fact that we are using Planck 2018 data and the information from all multipoles, while previous works used 2015 data and only low-$l$ data. 

We finish this section with a few remarks. The constraints on the stochastic GW spectrum derived in \cite{Namikawa:2014lla,Namikawa:2019tax}, and that we use to derive our bounds, assume a monochromatic top-hat GW spectrum at each momentum. In reality, our spectrum is broader so we expect that a dedicated CMB analysis could lead to even stronger constraints. On the other end, in \cite{Namikawa:2014lla,Namikawa:2019tax}  it is assumed that the stochastic GW spectrum is present at all times relevant for the CMB, from recombination until today. 
In the case of the DW network, the GW is mostly generated at the annihilation time, which can be much after recombination. Therefore, even though we do not expect this difference to have a large quantitative impact, it motivates a dedicated CMB analysis.

\section{Conclusion} \label{sec: Conclusion}

The recent measurement of a non-vanishing isotropic cosmic birefringence signal in the CMB,  hints towards new physics, potentially connected to axions.    
In this work, we have argued that networks of axionic defects that couple to photons, including the DWs studied in \cite{Takahashi:2020tqv,Kitajima:2022jzz} but also string and string-wall networks, can yield an isotropic birefringence signal in the ballpark of the recently measured signal.  In these scenarios,  $\mathcal{O}(2\pi f)$ field excursions are guaranteed  independently of initial conditions, and we argued that a nonzero isotropic signal is generic provided the network of defects is present at the last scattering surface (LSS).
We have also shown, using as a proxy dedicated 3D numerical simulations of domain wall networks, that the presence of defects comes along with a set of other signatures, both at the level of the birefringent signal and at the level of GWs, that can be probed with future data and that will help to disentangle the defect interpretation from other proposed explanations for the measurement. 

In  Sec. \ref{sec:CMB birefringence} we discussed the isotropic birefringence signals that are characteristic of the axionic defects.
We started by reviewing the case of DW networks and then argued that similar results apply to string and string-wall networks.  
Our reasoning was based on what we called \textit{environmental birefringence}: the fact that gradients and defects at low redshifts can induce large field excursions for the axion at large angles in the sky that, contrary to their high redshift partners, do not average out thus leaving an isotropic birefringence signal of environmental type. 
Compared to previous work, we identified a few more contributions to the isotropic signal, in the form of local gradients, besides the string loop contribution studied in \cite{Agrawal:2019lkr,Jain:2021shf,Jain:2022jrp}. These additional contributions can potentially overcome the constraints from the anisotropic birefringence spectrum and, therefore, it would be interesting to verify these statements with a dedicated simulation of axionic strings.

We then looked into the anisotropic part of the birefringence signal in Sec. \ref{sec:Anisotropic birefringence}.  The inhomogeneities of the network, peaked at the horizon scale, are a smoking gun for this scenario. 
We computed the anisotropic birefringence signal for a network that is in scaling at recombination and/or reionization using for the first time a dedicated 3D numerical simulation (with thermal and white noise initial conditions) of a domain wall network and the full-sky power spectrum, without the flat-sky approximation.  
We found that the reionization component is in slight tension with the current data from BICEP/Keck \cite{BICEPKeck:2022kci}, but a more detailed analysis should be done which takes into account the actual shape of the DW spectrum, for which we provide numerical fits in the Appendix \ref{app:FittingCl}. We have also estimated the effects on the anisotropic spectrum of the finite thickness of the LSS and the relativistic motion of the network and found them to be negligible at multiples close to the peak. Similarly, the monopole signal is free from the wash-out from the finite thickness of the LSS that happens for axion dark matter models.
We stress again that our results are based on simulations of domain wall networks in a $Z_2$ model, not really an axionic model. However, concerning these points we expect qualitatively similar conclusions for the string and string wall cases as the scalar power spectrum does not differ much among those different regimes \cite{Hiramatsu:2012sc}.

Then in section \ref{sec:Tomography}, we discussed how a tomographic analysis, through the measurement of the $EB$ power spectrum at low $l$ with future CMB surveys as proposed in \cite{Sherwin:2021vgb,Namikawa:2023zux}, 
would allow to distinguish if the network was formed before or after recombination formation and/or if it annihilated before or after reionization. In the process, we also argued that even if the network annihilates before today there can be situations where the isotropic signal remains due, for example, from misaligned contributions to the potential. This would suppress the signal at low $l$ and so alleviate the currently mild tension with BICEP/KECK data.
We summarized the different regimes and predictions for isotropic and anisotropic birefringence in table \ref{table II}.

Finally, in section \ref{sec: GWs}, we studied the gravitational effects of networks with DWs on the CMB anisotropies, which are proportional to the tension of the walls and independent of the birefringence signals. 
We assumed, conservatively, that the GW spectrum is maximal at the time the network annihilates and so used the aforementioned numerical simulations in the scaling regime to compute the GW spectrum today (see Fig. \ref{fig:OmegaGWplot}). Previous works placed CMB bounds \cite{Namikawa:2019tax} on stochastic GW backgrounds such as that generated by the DW network.
We used those existing bounds to constraint the energy density of the network $\Omega_\text{dw}^\text{ann}$ (or equivalently the DW tension) as a function of the annihilation temperature of the network (see Fig. \ref{fig:Bounds on OmegaGW}). Our results are the first bounds on DW networks that annihilate after recombination and bridge an existing gap in the literature: previous analyses have analyzed networks that either do not annihilate \cite{Zeldovich:1974uw,Lazanu:2015fua,Sousa:2015cqa} or annihilate before recombination \cite{Ramberg:2022irf}.
We ended by showing the prospects for detection with future CMB B-mode experiments. A possible measurement of a DW-like GW spectrum on CMB scales combined with DW-like birefringence signatures would  make a strong case for the presence of DWs after the recombination epoch.

This work has left a few research directions that would be interesting to explore in the future. We list here a few of those ideas. 
We have shown interesting predictions at the level of the birefringence spectrum. Therefore, it would be timely to perform a dedicated analysis of the detectability of these effects with near-future experiments such as Simons Observatory, LiteBIRD and CMB-S4.

Regarding the gravitational impact of the network on the CMB, here we restricted our analysis to the CMB B spectrum. However, DW networks also  impact the remaining CMB correlators, in particular, through the generation of isocurvature perturbations. 
It would be interesting to study these effects in more detail and, if possible, implement the backreaction effects of the network on the CMB perturbations directly on a CMB Boltzmann solver, in the spirit of what has been done in \cite{Lazanu:2015fua,Sousa:2015cqa}. 
That would allow us to improve the sensitivity to the tension of the wall and to perform further tests of the DW interpretation of the cosmic birefringence signal.

\section*{Acknowledgements}
We would like to thank Lara Sousa and Fabrizio Rompineve for discussions.
The work of RZF has been partly supported by the Direcci\'o General de Recerca del Departament d’Empresa i Coneixement (DGR), by the EC through the program Marie Sk\l odowska-Curie COFUND (GA 801370)-Beatriu de Pin\'os and by the FCT grants ~CERN/FIS-PAR/0027/2021, UIDP/04564/2020 and UIDB/04564/2020. The work of TH has been partly supported by JSPS KAKENHI Grant No. JP21K03559 and No. JP23H00110.  SG has the support of the predoctoral program AGAUR FI SDUR 2022 from the Departament de Recerca i Universitats from Generalitat de Catalunya and the European Social Plus Fund.
IO has the support of JSPS KAKENHI Grant No. JP20H05859 and No. 19K14702.
SG and OP acknowledge the support from the Departament de Recerca i Universitats from Generalitat de Catalunya to the Grup de Recerca `Grup de F\'isica Te\`orica UAB/IFAE' (Codi: 2021 SGR 00649) and the Spanish Ministry of Science and Innovation (PID2020-115845GB-I00/AEI/10.13039/501100011033). IFAE is partially funded by the CERCA program of the Generalitat de Catalunya.

\appendix

\section{Numerical Simulation}
\label{app: Numerical Simulation}

\subsection{Evolution equations and initial conditions}
\label{subapp:evolution_and_initial}

The field theory model that we simulate in this paper is described by the following action for a scalar field,
\begin{align}
S = -\int\!d^4x\,\sqrt{-g}\left[\frac{1}{2}\partial^\mu\phi\partial_\mu\phi + \frac{\lambda}{4}(\phi^2-v^2)^2\right],
\label{eq:action}
\end{align}
in the cosmological background,
\begin{align}
  ds^2 = a^2\left\{-d\eta^2 + (\delta_{ij}+h_{ij})dx^idx^j\right\},
\end{align}
where $\eta$ is the conformal time, $h_{ij}$ is the tensor perturbation.
Throughout this paper, we fix $\lambda=1$.
The action (\ref{eq:action}) yields the evolution equation of the scalar field,
\begin{align}
 \ddot{\phi} + 2\mathcal{H}\dot{\phi} - \partial^i\partial_i\phi = -a^2\lambda(\phi^2-v^2)\phi,
\end{align}
where the dot represents the derivative with respect to the conformal time and $\mathcal{H}:=aH$ is the conformal Hubble parameter. Defining the dimensionless variables, $\hat{\eta}:=v\eta$, $\hat{\boldsymbol{x}}:=v\boldsymbol{x}$, and $\psi:=a \phi/v$, we obtain
\begin{align}
 \ddot{\psi} -\frac{\ddot{a}}{a}\psi - \partial^i\partial_i\psi = -\lambda(\psi^2-v^2a^2)\psi.
 \label{eq:evol_scalar}
\end{align}
Henceforth, we omit the hat for the dimensionless variables and all physical quantities shall be normalised by $v$.
The action (\ref{eq:action}) yields also the evolution equations of the tensor perturbations,
\begin{align}
 \ddot{h}_{ij} + 2\mathcal{H}\dot{h}_{ij} - \partial^k\partial_k h_{ij} = \frac{2}{M_{\rm pl}^2}T_{ij},
\end{align}
where $M_{\rm pl}:=(8\pi G)^{-1/2}$.
Introducing $\chi_{ij}:=ah_{ij}$ and $\epsilon_G:=v/M_{\rm pl}$, we obtain
\begin{align}
 \ddot{\chi}_{ij} -\frac{\ddot{a}}{a}\chi - \partial^k\partial_k \chi_{ij} = 2a\epsilon_{G}^2\frac{T_{ij}}{v^2}.
  \label{eq:evol_chi}
\end{align}
We fix $\epsilon_G=0.1$, but the normalisation of $\chi_{ij}$ is not important since in Sec.~\ref{sec: GWs} we define the spectral function, $S(k,\eta)$, which is independent to $v$.
$T_{ij}$ is the spatial part of the energy-momentum tensor,
\begin{align}
 T_{\mu\nu} = \partial_{\mu}\phi\partial_{\nu}\phi - g_{\mu\nu}\left\{\frac{1}{2}\partial^{\alpha}\phi\partial_{\alpha}\phi+\frac{\lambda}{4}(\phi^2-v^2)^2\right\}.
\end{align}
We consider only the relevant term that sources the gravitational waves, which is given as
\begin{align}
 \frac{T_{ij}}{v^2} \supset \frac{1}{v^2}\partial_{i}\phi\partial_{j}\phi = \frac{1}{a^2}\partial_i\psi\partial_j\psi.
\end{align}
When we evaluate the gravitational wave spectrum, the transverse-traceless projection drops the other terms in the energy-momentum tensor.

In our simulations, we solve Eqs.~(\ref{eq:evol_scalar}) and(\ref{eq:evol_chi}) with the periodic boundary condition in the comoving box. We performed simulations with $L=20$ and $80$. The physical box size at $\eta$ is given as $a(\eta)L$ where we set $a(\eta_{\rm in})=1$. The number of grids is fixed to be $N=1024^3$.
We employ the Leap-Frog method and the central finite differences for the spatial derivatives, which achieves the second order both in time and space. We assume a matter-dominant Universe in which the conformal Hubble parameter is $\mathcal{H}=2/\eta$. The initial conformal time is $\eta_{\rm in} = 2.24$ and thus the Hubble parameter at this time is $\mathcal{H}(\eta_{\rm in}) = 0.894$.

For the initial conditions, we consider two different cases. One is the thermal noise for $\psi$, so that the two-point correlation functions of $\phi$ and $\dot{\phi}$ are given in \cite{Yamaguchi:2002sh} with a temperature $T_{\rm in}=v$. 
The second choice is the Gaussian random noise with $\sqrt{\langle\psi^2\rangle} = 0.1$ and $\langle\psi\rangle=0$. As the power spectrum of the Gaussian random noise is given as $\mathcal{P}(k)\propto k^2$, the power on small scales becomes too large. To smooth the initial field distribution, we apply a cooling period before starting a simulation according to Ref.~\cite{Hindmarsh:2018wkp}. We solve a diffusive equation replacing the second time-derivative of the field equation, $\ddot{\psi}$, by the first time-derivative, $\dot{\psi}$,
\begin{equation}
 \dot{\psi} = \partial^i\partial_i\psi  -\lambda(\psi^2-1)\psi,
\end{equation}
for $\eta_{\rm in}-\eta_{\rm diff} \leq \eta\leq \eta_{\rm in}$ with $\eta_{\rm diff}=100\Delta\eta_{\rm diff}$ and $\Delta\eta_{\rm diff}=3\times 10^{-3}$.  During this process, we do not consider the cosmic expansion by setting $a=1$.
For $\eta\geq \eta_{\rm in}$, we turn on the cosmic expansion and solve Eqs.~(\ref{eq:evol_scalar}) and (\ref{eq:evol_chi}).
The cooling process reduces the power of the fluctuations on small scales for $k>k_{\rm diff}$ where $k_{\rm diff}=1/\sqrt{\eta_{\rm diff}}$. We confirmed that the resulting power spectrum at the simulation end remains unchanged up to $\sqrt{\langle\psi^2\rangle} = 10$. 

\subsection{Area parameter}
\label{subapp:area_parameter}

\begin{figure}[t]
	\centering
	\includegraphics[width=0.48\linewidth]{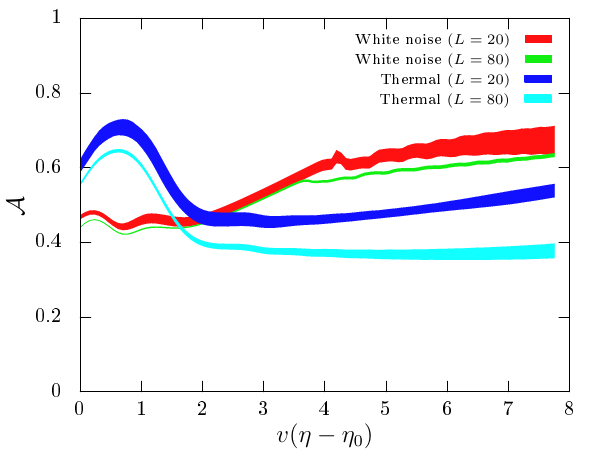} 
	\caption{Time evolution of the area parameter. The area parameter becomes almost constant in the late time, implying the scaling regime. We consider the cases with the white noise and thermal initial conditions in a small box ($L=20$) and a large box ($L=80$). Each result is averaged over 5 realisations, and the coloured bands indicate the $1\sigma$ error.}
	\label{fig:area_parameter}
\end{figure}

We use the algorithm introduced by Ref.~\cite{Press:1989yh} to estimate the comoving area of domain walls in the computational domain as
\begin{align}
 A = \Delta x^2\sum_{\rm links}\delta^{+}_{-}\frac{|\nabla\phi|}{|\partial\phi/\partial x|+|\partial\phi/\partial y|+|\partial\phi/\partial z|},
\end{align}
where $\Delta x$ is the lattice spacing and $\delta^{+}_{-}$ is defined to be 1 if a link between neighbour grid points crosses a wall and 0 if it does not. Then we introduce the area parameter defined as \cite{Hiramatsu:2012sc}
\begin{align*}
\mathcal{A} = \frac{\rho_{\rm dw}}{\sigma}t = \frac{2A}{3L^3\mathcal{H}},
\end{align*}
where $\rho_{\rm dw}$ and $\sigma$ are the energy density and the tension of domain walls, respectively, $A$ is the comoving area and $L$ is the comoving box size. 
Note that we find $c=3\mathcal{A}/2$ in eq.~(\ref{eq:rho_dw}). The comoving area of the domain wall is proportional to time, just as the correlation length of the cosmic string in the scaling regime is proportional to time. Therefore, the area parameter becomes constant in the scaling regime. In Fig.~\ref{fig:area_parameter}, we show the time evolution of the area parameter with $L=20, 80$ and the thermal and the white-noise initial conditions with the cooling process. Although there is a scatter, the area parameter becomes almost constant and converges to $\mathcal{A} \approx 0.5\pm 0.1$.

\section{Angular Power Spectrum}
\label{app:angular PS}
In this section, we review the computation of the anisotropic birefringence power spectrum (see e.g. \cite{Greco:2022ufo,Greco:2022xwj} for more details).  The coefficients of the multipole expansion in eq. \eqref{eq:betadeco} are given by
\begin{equation}
    \beta_{\ell m}(\eta)=\int{\rm d}^2\mathbf{n} \, Y^*_{\ell m}(\mathbf{\Hat{n}})\, \beta(\eta,\Delta\eta \, \mathbf{\Hat{n}}),
\end{equation}
with $\Delta\eta=(\eta_0-\eta)$ the comoving distance between the time $\eta$ and the present time $\eta_0$. By taking the Fourier transform of the field
\begin{equation}\label{eq:C1}
\beta(\eta,\Delta\eta \,\mathbf{\Hat{n}})=\int\frac{{\rm d}^3\mathbf{k}}{(2\pi)^3}e^{i\Delta\eta \, \mathbf{\Hat{n}\cdot\Hat{k} }}\beta(\eta,\mathbf{\Hat{k}})
\end{equation}
and using the plane wave expansion in terms of spherical harmonics
\begin{equation}
e^{i\Delta\eta \, \mathbf{\Hat{n}\cdot\Hat{k} }}=4\pi\sum_{\ell m}i^\ell j_\ell (k\Delta\eta)Y^*_{\ell m}(\mathbf{\Hat{k}})Y_{\ell m}(\mathbf{\Hat{n}}) \,
\end{equation}
where $j_\ell $ is the $\ell -$th spherical Bessel function, 
one can write eq. \eqref{eq:C1} as
\begin{equation}
    \beta_{\ell m}(\eta)=4\pi i^\ell\int\frac{{\rm d}^3\mathbf{k}}{(2\pi)^3}Y^*_{\ell m}(\mathbf{\Hat{k}})j_\ell(k\Delta\eta)(\mathbf{\Hat{k}}) \beta(\eta,\mathbf{\Hat{k}})
\end{equation}
where we have used the orthonormality property of the spherical harmonics 
\begin{eqnarray}
    \int {\rm d}^2\mathbf{\Hat{k}} \,Y^*_{\ell_1 m_1}(\mathbf{\Hat{k}})Y_{\ell_2 m_2}(\mathbf{\Hat{k}})=\delta_{\ell_1\ell_2}\delta_{m_1 m_2}. \label{eq:orthonormality spherical harmonics}
\end{eqnarray}
Then we can write the two-point function as
\begin{align}\label{eq:C6}
\langle \beta^*_{\ell_1 m_1}(\eta_1) \beta_{\ell_2 m_2}(\eta_2)\rangle = &(4\pi)^2i^{\ell_2-\ell_1}\int\frac{{\rm d}^3\mathbf{k}_1}{(2\pi)^3}\int\frac{{\rm d}^3\mathbf{k}_2}{(2\pi)^3}Y^*_{\ell_1 m_1}(\mathbf{\hat{k}}_1)Y^*_{\ell_2 m_2}(\mathbf{\hat{k}}_2) \nonumber \\
&\quad \times j_{\ell_1}(k_1\Delta\eta_1)j_{\ell_2}(k_2\Delta\eta_2)\langle \beta^*(\eta_1,\mathbf{\hat{k}}_1) \beta(\eta_2,\mathbf{\hat{k}}_2)\rangle.
\end{align}
Using eq. \eqref{eq:betadeco} and the definition of the scalar power spectrum \eqref{eq: Scalar field power spectrum}, this simplifies to
\begin{equation}\label{eq:theta PS}
    \langle \beta^*_{\ell_1 m_1}(\eta_1)\beta_{\ell_2 m_2}(\eta_2)\rangle = 4\pi \left(\frac{c_\gamma\alpha_\text{em}}{4\pi }\right)^2 \delta_{\ell_1\ell_2}\delta_{m_1 m_2}\int\frac{{\rm d}k }{2\pi^2}k^2j_{\ell_1}(k\Delta\eta_1)j_{\ell_2}(k\Delta\eta_2)P_{k}^\theta(\eta_1,\eta_2),
\end{equation}
Note that for the anisotropic part of the birefringence, we just care about the field value at emission since $\theta(\eta_0)$ in eq. \eqref{eq:betadeco} only contributes to the monopole $\ell=0$. By evaluating eq. \eqref{eq:theta PS} at recombination we get that
\begin{equation}\label{eq:PS rec}
    C^{\beta\beta}_\ell |_{\rm rec}= \frac{2}{\pi} \left(\frac{c_\gamma\alpha_\text{em}}{4\pi }\right)^2 \int{\rm d}kk^2j^2_{\ell}(k\Delta\eta_{\rm rec})P_{k}^\theta(\eta_{\rm rec}).
\end{equation}
We can make contact with the result of \cite{Kitajima:2022jzz} obtained in the flat-sky approximation by using the fact that for high $\ell$ the Bessel function $j_{\ell}(k\Delta\eta)$ is well approximated by a delta function which peaks at $\ell=k\Delta\eta$, therefore, we can write eq. \eqref{eq:PS rec} as
\begin{equation}\label{eq:PS approx}
    C^{\beta\beta}_\ell |_{\rm rec}\simeq  \frac{4\pi^2}{\pi} \left(\frac{c_\gamma\alpha_\text{em}}{4\pi }\right)^2 \left( \mathcal{P}^\theta_{k=\frac{\ell}{\Delta\eta_{\rm rec}}} \right) \int_0^\infty{\rm d}\ln{x} \, j^2_{\ell}(x)=4\pi\left(\frac{c_\gamma\alpha_\text{em}}{4\pi }\right)^2 \left(\mathcal{P}^\theta_{k=\frac{\ell}{\Delta\eta_{\rm rec}}} \right)\frac{1}{2\ell(1+\ell)}
\end{equation}
where we have assumed that $P^\theta_k$  is approximately constant in the region where the Bessel function has support, used the well-known result $\int_0^\infty{\rm d}\ln{x} \, j^2_{\ell}(x)=1/{2\ell(1+\ell)}$ \cite{Baumann} and defined the adimensional quantity $\mathcal{P}^\theta_k=\frac{k^3}{2\pi^2}P^\theta_k$. Then using the fact that $\beta\simeq c_\gamma \alpha_{\rm em}/4$ and $\Delta\eta_{\rm rec}\simeq \eta_0$ we find 
\begin{equation}\label{eq:Flatskyres}
    \frac{\ell(\ell+1)}{\beta^2 }C^{\beta\beta}_\ell\simeq\frac{2}{\pi}\mathcal{P}^\theta({\ell}/{\eta_{0}}),
\end{equation}
this is analogous to the formula given in \cite{Kitajima:2022jzz}, but in terms of the 3D power spectrum. In Figure \ref{fig:Anisotropic birefringence}, we show the difference between this flat-sky approximation and the full-sky result. 

\section{Fitting the $\mathcal{C}_\ell$}
\label{app:FittingCl}
\begin{figure}
	\centering
	\includegraphics[width=0.48\linewidth]{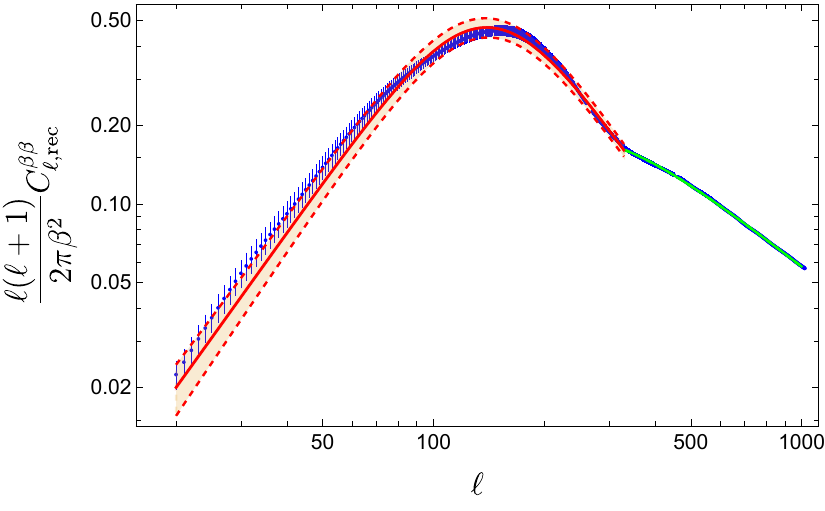} \, \,
		\includegraphics[width=0.48\linewidth]{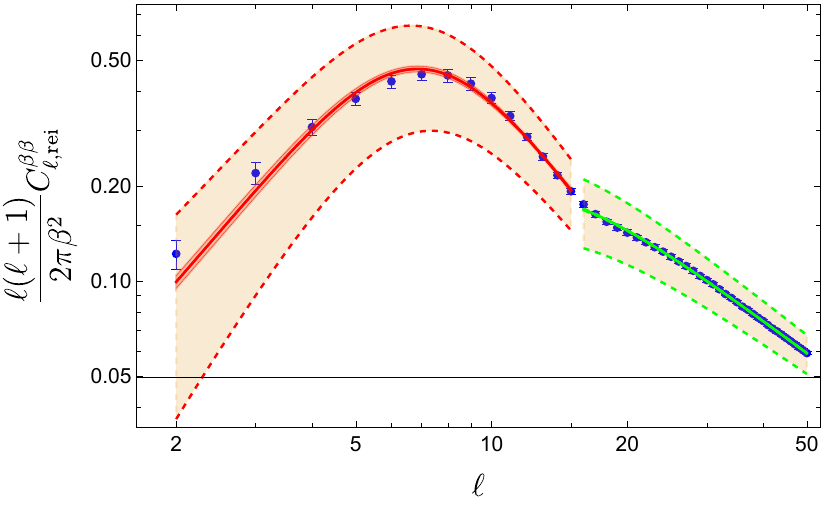}
	\caption{The left panel shows the anisotropic spectrum of birefringence of recombination whereas the right panel shows it of reionization. In both panels, we show the numerical data points with the respective errors coming from 5 realizations in blue. In green and red we show the numerical fits which describe the two power-law behaviours at multiples higher than $\ell_{peak}$ as explained in the text. The error band between the dashed lines comes from the cosmic variance.  }
	\label{fig:Clfits}
\end{figure}

In this section, we provide numerical fits for the spectra of the anisotropic birefringence that could be useful in future searches. We found our numerical results to be better described, at multipoles higher than the peak, by a change in the power law behaviour as can be seen in Figure \ref{fig:Clfits}. In both regimes, we use the following function to fit the $C^{\beta\beta}_\ell$  
\begin{equation}
f(x)= \frac{A}{1+(\frac{x}{x_{peak}})^\delta}.
\end{equation}
In the case of recombination the change of behaviour is found around  $\ell\simeq 330$ and the best-fit parameters are $A=0.0003$, $x_{peak}=140$, $\delta=4$ for $\ell<330$ and $A=0.00005$, $x_{peak}=201$, $\delta=3$ for $\ell>330$. Note that in Figure \ref{fig:Clfits} we plot  $\ell(\ell+1)C^{\beta\beta}_\ell / 2\pi\beta^2$ with $\beta=c_\gamma \alpha_{\rm em}/4$ such that the error associated with $\beta$ is factorized out.  At low multiples $\ell<\ell_{peak}$, the spectrum goes like $\ell(\ell+1) C^{\beta\beta}_\ell\propto \ell^2$. Close to the peak at  $\ell>\ell_{peak}$, it has a steeper slope $\ell(\ell+1) C^{\beta\beta}_\ell\propto \ell^{-2}$ whereas at even higher multiples $\ell(\ell+1) C^{\beta\beta}_\ell\propto \ell^{-1}$ which is compatible with the scaling of the power spectrum $\mathcal{P}_\phi\propto (k\eta)^{-1.3}$ that we discussed in section \ref{sec:Anisotropic birefringence}. For reionization, the best-fit parameters are $A=0.1$, $x_{peak}=7$, $\delta=3.9$ for $\ell<16$ and $A=0.02$, $x_{peak}=9.7$, $\delta=3$ for $\ell>16$.  The bands enclosed between the dashed lines show the error from cosmic variance approximated as $\Delta C_\ell^{\beta\beta}\simeq \sqrt{2/2\ell+1} C_\ell^{\beta\beta}$ \cite{Takahashi:2020tqv}. As it's evident from  Figure \ref{fig:Clfits}, the cosmic variance error is much more important at lower multiples being the dominant source of error for the spectrum of reionization whereas it's of the same order as the numerical one in the case of recombination.

\section{Fitting the GW spectrum}
\label{app: Fitting the GW spectrum}

In this section, we provide more details on how we fitted the GW spectrum from the numerical simulations. We focus on the cases where the simulation box has size $N=1024$ and $L=80$ (see App. \ref{app: Numerical Simulation}).

\begin{figure}
	\centering
	\includegraphics[width=0.48\linewidth]{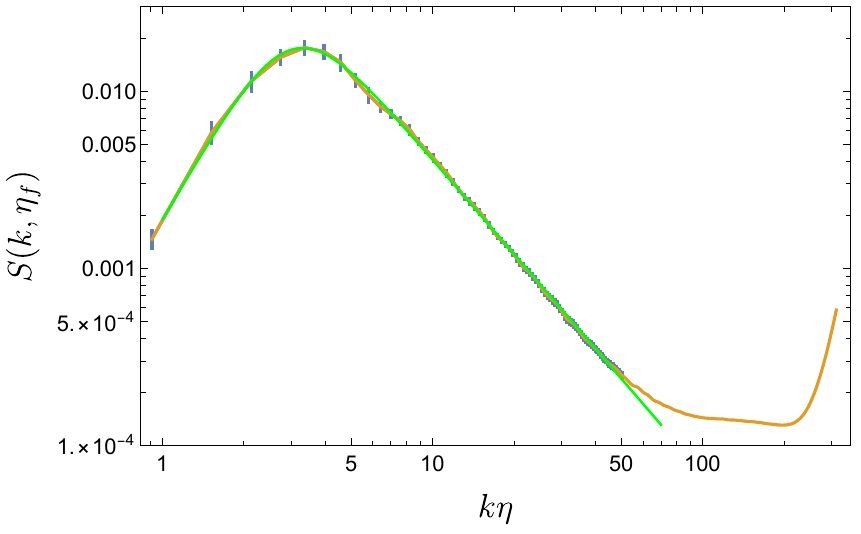} \, \,
		\includegraphics[width=0.48\linewidth]{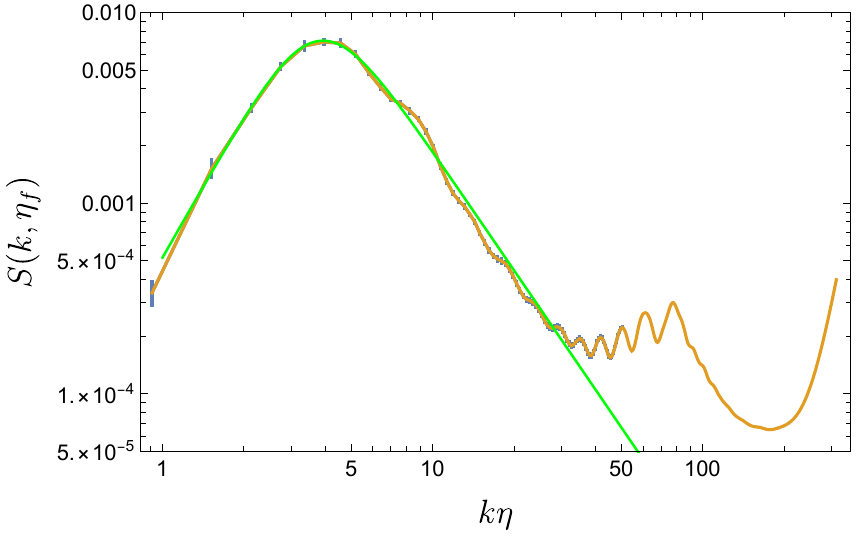}
	\caption{In blue are the data points used to perform the fit and their 1-sigma errors. The orange curve shows the full output of the simulation and the green curve the fitting function. Left plot: thermal initial conditions, Right plot: White Noise initial conditions.}
	\label{fig:scaledgwspectrum-error-thermal-l80}
\end{figure}

As explained in the main text, to perform the fit we used the simulation data at the latest time $\eta_f$, when the spectrum has approximately converged to an attractor shape. 
We show in Fig.\ref{fig:scaledgwspectrum-error-thermal-l80} in orange the full spectrum at $\eta_f$. It is clear that there is a change in slope at the small scale $k\eta \sim 50$ in both the simulations with thermal and white noise initial conditions. Therefore, for the purpose of performing the fit we discarded the data at $k\eta > 50$ as these scales are more sensitive to the finite width of the DW which for numerical reasons cannot be much smaller than the Hubble radius $H^{-1}$. However, we do not discard the possibility that the change in slope is physical. A dedicated study of this effect using larger numerical simulations would be needed to confirm this aspect. 

We show in Fig. \ref{fig:scaledgwspectrum-error-thermal-l80} in blue the data points that were used to perform the fits, and their $1\sigma$ errors obtained after averaging the simulation over 5 realizations. 
We have then used the  {\tt NonlinearModelFit} function of {\tt Mathematica} \cite{Mathematica} that performs a nonlinear regression of the formula in eq. \ref{eq: Fitting formula} to the data points including the $1\sigma$ errors. The best-fit values and standard errors for the parameters are shown in Table 2.

We note however that when including all the data in the simulation (apart from the last growing part of the spectrum which is a clear numerical artifact) we find that the best fit value for the slope of the spectrum in the UV gets closer $\theta_1$ which is the value obtained in previous studies \cite{Hiramatsu:2013qaa}.

\vspace{0.2cm}

\begin{table}
	\begin{center}
		\begin{multicols}{2}
			\begin{tabular}{l|ll}
				\text{} & \text{Estimate} & \text{Standard Error} \\
				\hline
				A & 0.0175 & 0.0007 \\
				$\beta$ & 2.6 & 0.1 \\
				$\theta$ & 1.772 & 0.008 \\
				$x_{\text{peak}}$ & 3.30 & 0.06
			\end{tabular} \\
						\begin{tabular}{l|ll}
				\text{} & \text{Estimate} & \text{Standard Error} \\
				\hline
				A & 0.0071& 0.00007 \\
				$\beta$ & 2.49 & 0.09 \\
				$\theta$ & 2.07 & 0.04 \\
				$x_{\text{peak}}$ & 3.93 & 0.03
			\end{tabular} 
		\end{multicols}
		\caption{Best-fit values for thermal (left) and white noise (right) initial conditions.}
     \label{table: best fit values}
	\end{center}
\end{table}

\bibliography{biblio}
\end{document}